\documentclass[12pt,draftclsnofoot,journal,onecolumn]{IEEEtran}

\usepackage{graphicx} 		
\usepackage{subfigure}
\usepackage{amssymb} 		
\usepackage{algorithm}  
\usepackage{algorithmic}
\usepackage{footnote}        	
\usepackage{amsmath}      	
\usepackage{cite}
\usepackage{cases}
\usepackage{multirow}
\usepackage{color}
\usepackage{url}

\IEEEoverridecommandlockouts

\newtheorem{lemma}{Lemma}

\newtheorem{optimization}{Robust Optimization Problem}

\begin{document}
\title{Robust Cooperative Spectrum Sensing Scheduling Optimization in Multi-Channel Dynamic Spectrum Access Networks}
\author{Chun-Hao Liu,~\IEEEmembership{Student Member,~IEEE}, Arash Azarfar,~\IEEEmembership{Student Member,~IEEE}, Jean-Fran\c{c}ois Frigon,~\IEEEmembership{Senior Member,~IEEE}, Brunilde Sans\`o,~\IEEEmembership{Member,~IEEE}, and Danijela Cabric,~\IEEEmembership{Senior Member,~IEEE}%
\IEEEcompsocitemizethanks{\IEEEcompsocthanksitem Chun-Hao Liu and Danijela Cabric are with the Department of Electrical Engineering, University of California, Los Angeles, 56-125B Engineering IV Building, Los Angeles, CA 90095-1594, USA (email: \{liuch37, danijela\}@ee.ucla.edu).
\IEEEcompsocthanksitem Arash Azarfar, Jean-Fran\c{c}ois Frigon, and Brunilde Sans\`o are with the Department of Electrical Engineering, {\'E}cole Polytechnique de Montr{\'e}al, C.P. 6079, succ. centre-ville, Montr{\'e}al, QC, Canada, H3C 3A7 (email: \{arash.azarfar, j-f.frigon, brunilde.sanso\}@polymtl.ca).
\IEEEcompsocthanksitem This research was supported by the National Science Foundation under CNS grant 1149981, NSERC under Grant STPG365205, and FQRNT-CREER program for international internship application No. 179776.
\IEEEcompsocthanksitem Part of this work is presented in IEEE GLOBECOM, 2014~\cite{Arash14}.}}

%
%
%

\IEEEcompsoctitleabstractindextext{
\begin{abstract} 
Dynamic spectrum access (DSA) enables secondary networks to find and efficiently exploit spectrum opportunities. A key factor to design a DSA network is the spectrum sensing algorithms for multiple channels with multiple users. Multi-user cooperative channel sensing reduces the sensing time, and thus it increases transmission throughput. However, in a multi-channel system, the problem becomes more complex since the benefits of assigning users to sense channels in parallel must also be considered. A sensing schedule, indicating to each user the channel that it should sense at different sensing moments, must be thus created to optimize system performance. In this paper, we formulate the general sensing scheduling optimization problem and then propose several sensing strategies to schedule the users according to network parameters with homogeneous sensors. Later on we extend the results to heterogeneous sensors and propose a robust scheduling design when we have traffic and channel uncertainty. We propose three sensing strategies, and, within each one of them, several solutions, striking a balance between throughput performance and computational complexity, are proposed. In addition, we show that a sequential channel sensing strategy is the one to be preferred when the sensing time is small, the number of channels is large, and the number of users is small. For all the other cases, a parallel channel sensing strategy is recommended in terms of throughput performance. We also show that a proposed hybrid sequential-parallel channel sensing strategy achieves the best performance in all scenarios at the cost of extra memory and computation complexity.
\end{abstract}

}

\maketitle

\IEEEdisplaynotcompsoctitleabstractindextext

%
\IEEEpeerreviewmaketitle

\section{Introduction}
\label{sec:intro}
In order to increase current spectrum utilization, it has been proposed that secondary (unlicensed) users (SUs) could efficiently exploit spectrum vacancies that are normally licensed to primary users (PUs) in either temporal, frequency, or spatial domain in dynamic spectrum access (DSA) networks~\cite{zhao07}. The two-stage sense and transmit processing is a well-known DSA strategy for SUs~\cite{Hoang08}. SUs first sense licensed channels and, if a channel is not occupied by PUs, the SUs can then transmit on it. Otherwise, the SUs need to sense other channels to find possible transmission. The sensing strategy is important for the performance of the SUs since if licensed channels are sensed in a shorter time, SUs will have a longer access opportunity on the vacant channels, which results in a larger throughput. 

The sensing strategy over multiple channels for SUs is therefore an important issue that needs to be addressed. That strategy, designed to optimize an objective metric, such as throughput, delay or energy consumption, provides a time schedule to sense channels for SUs, so that a decision is made as at which time instant each SU should sense which channels. If multiple SUs are assigned to sense the same channel at the same time, a cooperative sensing is then pursued for this channel, which can increase sensing accuracy and may reduce sensing time~\cite{Brodersen06}.
\subsection{Related Work}
\label{sec:related_work}
Sensing strategies have so far been mostly investigated in what relates to sensing order optimization and acquiring the stopping time in a sequential manner where channels are sensed one after the other. To the best of our knowledge,~\cite{Shin08} is the first to introduce the concept of sensing order. The authors proposed to sense channels in the decreasing order of the probability of being idle.~\cite{liu13} also takes channel capacity and sensing time into account to derive the optimal sensing order. A multi-user network is investigated in~\cite{azarfar13} where channels are being sensed in parallel, but the only parameter used for the decision making is channel occupancy without considering the impact of cooperative sensing. In~\cite{Datla09}, the authors proposed a scheduling scheme for spectrum sensing based on the idea that when a channel is free, the channel can be sensed with a lower time resolution set based on a backoff scheme.~\cite{Giannakis08} proposed a robust routing schedule to maximize the social network utility subject to the variance constraint.~\cite{Cui14} proposed an online decision scheduling algorithm to determine the sensing period together with a sequential detection for spectrum sensing, which is robust to short-term channel change and possible data outliers. 

\subsection{Our Contribution}
\label{sec:contribution}
In addition to the sequential sensing order optimization which is analyzed considering all physical layer details, unlike previous works, this paper is also the first which formulates the general sensing strategy problem and addresses the compromises that exists between parallel and sequential sensing strategies, i.e., assigning less users to each channel in order to sense a large number of channels in parallel versus the benefits of assigning multiple users to each channel to cooperatively sense the same channel. Therefore, we propose several structured sensing strategies to maximize system throughput, and we investigate the tradeoff among these strategies under various circumstances. Finally, we discuss the robust design when the proposed sensing strategies encounter uncertainty in PU channel occupancy and detection signal-to-noise ratio (SNR). The contribution of this paper is thus threefold:
\begin{enumerate}
\item We introduce and formulate the general problem of sensing strategy for optimal sensing allocation of SUs to maximize system throughput. However, due to implementation and analysis complexity, the general problem is not solved. This is one of the limitation of this work; 
\item Three classes of structured sensing strategies, i.e., sequential\footnote{The optimal sequential strategy has been proposed in our previous work~\cite{liu13}. But in~\cite{liu13}, we assume arbitrary sensing time. Here we consider the practical physical layer sensing method to obtain the sensing time considering user cooperation.}, parallel, and sequential-parallel multi-channel sensing strategies are proposed, resolved with optimal and heuristic algorithms, and compared in presence of homogeneous and heterogeneous sensors;
\item A robust optimization for the proposed strategies is provided to investigate how the sensing strategy decision is affected when there is uncertainty for the detected PU SNR and channel occupancy. 
\end{enumerate}
The reminder of the paper is organized as follows. The system model and problem formulation are provided in Section~\ref{sec:system_model}. The general and all the particular sensing strategies are presented and analyzed in Section~\ref{sec:strategy}. Section~\ref{sec:hetsensors} investigates the case of heterogeneous sensors where sensing SNRs are different for different sensors and channels. In Section~\ref{sec:robust_scheduling}, the sensing strategies are analyzed in the presence of uncertainty, and a robust optimization is provided. Numerical results are provided and discussed in Section~\ref{sec:numerical_results}. Finally, Section~\ref{sec:conclusions} concludes the paper.
\section{System Model and Problem Formulation}
\label{sec:system_model}
\begin{figure}
\centering
\includegraphics[width=0.68\columnwidth]{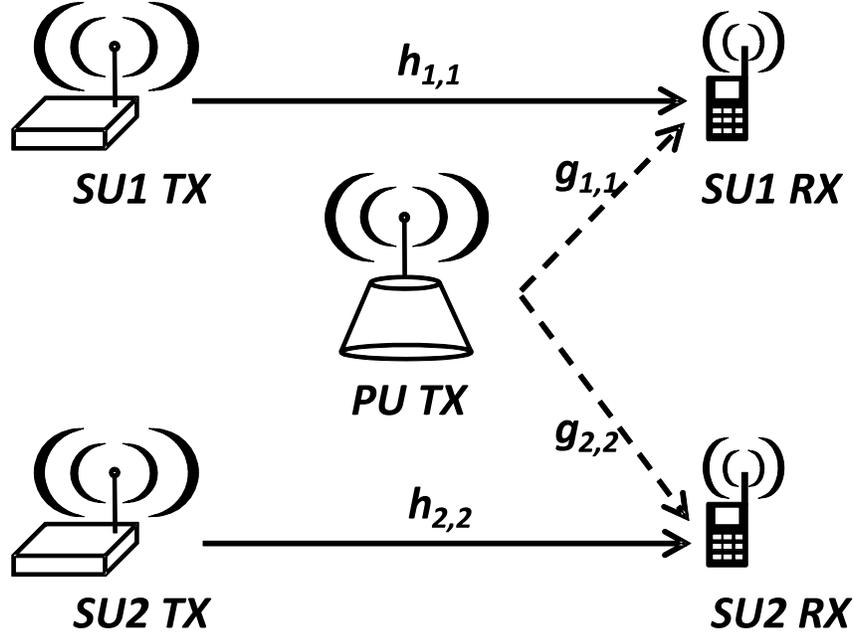}
\caption{The network model with a primary user transmitter (TX), two SU TXs, and two SU receivers (RXs).}
\label{Fig:Network_Model}
\end{figure}

We consider a DSA network with $N$ SUs transceiver pairs and $M$ channels as shown in Fig.~\ref{Fig:Network_Model}. Similar to most of the works in the literature, channels and PUs' activity in the channels are assumed to be fully independent~\cite{Shin08}. 
PUs are assumed to transmit synchronously on the channels in a time-slotted fashion with a slot duration equal to $T$~\cite{liu13}. Note that the time slot length in our work is the period during which the channel and the traffic statistics can be considered almost invariant. At the beginning of each time slot, the SU central network controller determines the sensing strategy for SUs to maximize the total expected spectrum opportunities for transmission. A spectrum sensing strategy includes the time schedule (e.g., sensing order) and job schedule (which users sense which channels). After the users finish sensing a channel, sensing results are sent to the central controller where they will be merged to make the final scheduling decision. Since our work mostly focuses on the sensing scheduling aspects of the problem, the transmission delay of sensing results to the controller is not considered\footnote{In the literature, some works addressed this issue. For instance, constraints on the number of reporting sensors is discussed in~\cite{Digham12}.}.  

The channel gain between the $i$-th SU transceiver pair operating on the $m$-th channel is denoted as $h_{m,i}$, and the channel gain from PU transmitter to the $i$-th SU receiver operating on the $m$-th channel is denoted as $g_{m,i}$. We thus define the $i$-th SU transmission capacity on the $m$-th channel as $C_{m,i}=B_m\log_2(1+\Gamma_{m,i})$, where $B_m$ is the bandwidth of the $m$-th channel,~$1\leq m\leq M$, $\Gamma_{m,i}=\frac{P_i h_{m,i}^2}{\sigma_m^2}$ is the received SNR for the $i$-th SU on the $m$-th channel, $P_i$ is the transmission power for the $i$-th SU, and $\sigma_m^2$ is the noise power on the $m$-th channel. For simplicity, we assume $\Gamma_{m,i}$ are the same for all SUs, which will be reduced to $\Gamma_m$ and hence $C_m$. For heterogeneous sensors, we define the corresponding detection SNR as $\gamma_{m,i}=\frac{\Psi_{m} g_{m,i}^2}{\sigma_m^2}$, where $\Psi_m$ is the transmission power of the primary user at channel $m$ and $g_{m,i}$ is the channel gain from PU to the $i$-th SU receiver on the $m$-th channel.
Note that for simplicity we assumed channel sensing is performed at the SU receiver node. For the homogeneous case, the detection SNR is denoted by $\gamma$. The probability for the $m$-th channel of being occupied by primary users is assumed to be known at the central controller as $u_m$, where $u_m$ can be estimated or measured efficiently~\cite{liu13,Casadevall13} in the training phase, as will be discussed in Section \ref{sec:robust_scheduling}. 
\subsection{Cooperative Spectrum Sensing}
\label{sec:cooperative_sensing}
Cooperative sensing is a well-known solution to enhance sensing performance~\cite{ghasemi05}. The reason is that as the collective decision is made with several individual sensing results, the requirement of sensing accuracy for each individual user can be lowered, hence the sensing time can be reduced. In a time slotted DSA network, since the status of the channel does not change during one time slot, minimizing the sensing time for the channel implies increasing the expected transmission throughput~\cite{Hoang08} for SUs.   
\subsubsection{Primary User Detection}
\label{sec:energy_detection}
Consider a secondary receiver that needs to detect primary users sending pilot signals on a particular channel~\cite{Brodersen06}. Let $\tau$ be the sensing time and assume that the receiver's sampling frequency is $f_s$ such that $N_s=\tau\times f_s$ samples are gathered to make the decision of whether a channel is occupied by a primary user. The minimum sensing time required to satisfy the given detection quality under additive white Gaussian noise (AWGN) channel by the optimal detector, i.e., the matched filter is equal to:
\begin{equation}
\tau=\frac{\left[Q^{-1}(P_f)-Q^{-1}(P_d)\right]^2}{\gamma f_s},
\label{eq:sensing_time}
\end{equation}
where $\gamma$ is the detected SNR and $P_d$ is the probability of detection, defined as the probability of detecting the primary user when it is present. $P_f$ is the probability of false-alarm defined as the probability of wrongly finding the channel occupied when it is actually vacant.  Note that even though we consider only AWGN channels, the discussion can be extended to any detection model as long as $P_d$ and $P_f$ are represented as a function of SNR. For example, $P_d$ and $P_f$ as a function of fading parameters can be found in~\cite[Section V]{Hoang08}. In addition, we choose the sensing sampling frequency as the Nyquist frequency which equals to two times the corresponding channel bandwidth in our simulations. 
\subsubsection{Fusion Rules}
\label{sec:fusion}
Sensing results reported by different users may be combined in different manners, known as fusion rules~\cite{ghasemi05}. In what follows, we discuss~\emph{OR} and~\emph{AND} hard fusion rules because they are commonly used in the literature and also they provide bounds for the more general rule~\emph{$k$-out-of-$N$}. Assume all $N$ users are homogeneous, i.e., they have the same $P_d$ and $P_f$. Thus, the cumulative probability of detection and false-alarm are given as
$Q_d=1-(1-P_d)^N$, and $Q_f=1-(1-P_f)^N$ for the~\emph{OR} fusion rule respectively, and as $Q_d=P_d^N$, and $Q_f=P_f^N$ for the~\emph{AND} fusion rule respectively. 

\begin{figure}%
\centering
\includegraphics[width=0.68\columnwidth]{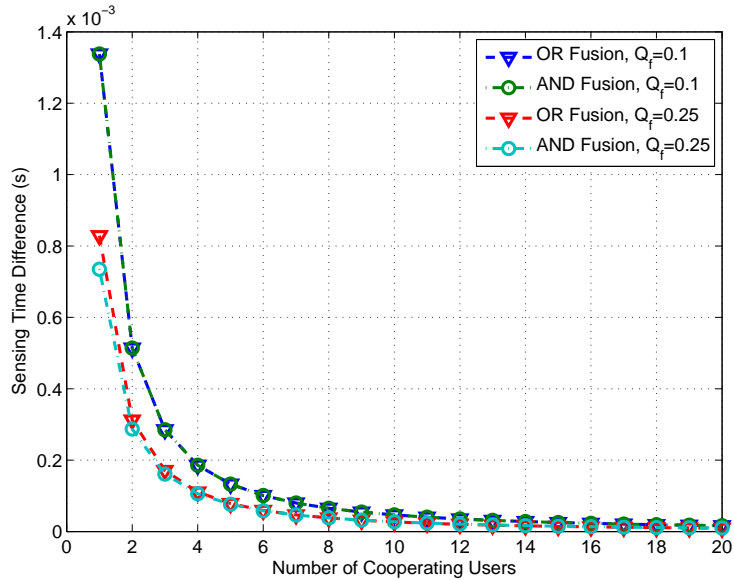}%
\caption{The gain of incorporating more users into the cooperative sensing process decreases when the number of users increases. The used parameters are $f_s=5$ kHz, $Q_d=0.9$, and SNR is $-5$ dB.}%
\label{GainVsUserIncrease.pdf}%
\end{figure}

From equation~(\ref{eq:sensing_time}), the minimum cooperative sensing time by $N$ homogeneous users to satisfy the $Q_d$ and $Q_f$ is expressed as
\begin{align}
\tau=\begin{cases}
\frac{\left[Q^{-1}(1-\sqrt[N]{1-Q_f})-Q^{-1}(1-\sqrt[N]{1-Q_d})\right]^2}{\gamma f_s} & \!\!\!\! \text{for the~\emph{OR} Rule},\\
\frac{\left[Q^{-1}(\sqrt[N]{Q_f})-Q^{-1}(\sqrt[N]{Q_d})\right]^2}{\gamma f_s} & \!\!\!\! \text{for the~\emph{AND} Rule}.
\label{eq:Coop-sensing_time}
\end{cases}
\end{align}
Throughout the paper, we also define $\tau_{m,n}$ as the cooperative sensing time of channel $m$ by $n$ sensors. Equation~(\ref{eq:Coop-sensing_time}) provides two important insights. First, for any channel $m$ (we thus drop the channel index), $\tau_{n}$ is a decreasing function of $n$. The other insight is related to the number of cooperative sensors. As illustrated in Fig.~\ref{GainVsUserIncrease.pdf}, the sensing time gain $\tau_n-\tau_{n+1}$, i.e., adding another user to the process of cooperative sensing, decreases when $n$ increases. The most improvement in cooperative sensing time is therefore obtained when two users cooperate instead of sensing a channel by one user. This behavior promotes the idea of distributing the users more evenly among channels. We will use this result in Section~\ref{subsec:analytical-discuss}. 
Note that in the discussions above, it is assumed that the detection SNR $\gamma$ is given and the same for all users. The case with heterogeneous sensors (different detection SNRs thus $P_d$ and $P_f$ for different users) and the case with a random SNR will be investigated in Section~\ref{sec:hetsensors} and~\ref{sec:robust_scheduling}, respectively. 
%
\subsection{Problem Formulation}
\label{sec:problem}
We define the beginning of a time slot as the reference point $t=0$, and the elapsed time when the sensing process for channel $i$ is finished as $T^{(i)}_{I}$. Note that $T^{(i)}_{I}$ depends on the sensing order and user allocation schemes, and the sensing time for channel $i$ depends on the number of users allocated to it. As illustrated in Fig.~\ref{Fig:StrategiesExample}(a), if channel $i$ is found available, it is a potential spectrum opportunity with duration $T-T^{(i)}_{I}$. The expected throughput obtainable from the spectrum opportunity of channel $i$ is thus equal to $C_i(1-u_i)(T-T^{(i)}_{I})$. The elapsed time for a channel which is not sensed can be assumed to be $T$ (no throughput gain). Our objective is to maximize the total expected normalized throughput $R$ from all channel spectrum opportunities\footnote{In this paper, our focus is on the spectrum opportunity detection part and our objective is thus to maximize the potential throughput for other transmitting users which do not participate in the sensing process. The transmission scheduling problem where users participate both in sensing and transmission is out of the scope of this paper and remains as our future work. The potential throughput in~(\ref{eq:Throughput}) thus represents an upper bound on the actual network throughput when joint sensing and transmission assignment of users is taken into account.} by deciding the optimal sensing strategy, i.e.,
\begin{align}
\max\limits_{\mathcal{A}} {\mathbb{E}\{R(\mathcal{A})\}=\sum_{i=1}^{M}{\frac{(T-T_I^{(i)}(\mathcal{A}))C_{i}(1-u_{i})}{T}}},
\label{eq:Throughput}
\end{align} 
where $\mathbb{E}\{\cdot\}$ is the expectation operation and $\mathcal{A}$ is a sensing strategy. Note for any channel $i$, $T_I^{(i)}(.)$ is a function of $\mathcal{A}$. 
\section{Spectrum Sensing Strategies with Homogeneous Sensors}
\label{sec:strategy}
\begin{figure}
\centering
\includegraphics[width=0.68\columnwidth]{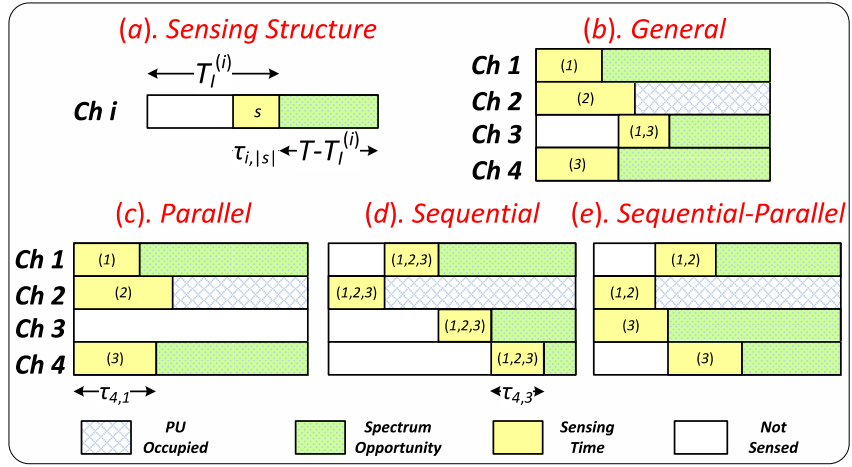}
\caption{(a) Sensing structure when channel $i$ is sensed by a subset $s$ with $|s|$ users. (b) A general sensing strategy where user $1$ waits for user $3$ to finish its job and then they sense channel $3$ cooperatively. (c) An example of parallel strategy where channel $3$ is not sensed. (d) An example of sequential strategy with the channel (Ch) sensing order (Ch $2$,Ch $1$,Ch $3$,Ch $4$) by all users. (e) Sequential-parallel strategy where (Ch $2$, Ch $1$) are sensed sequentially by users $1$ and $2$ and in parallel for (Ch $3$, Ch $4$) by user $3$.}
\label{Fig:StrategiesExample}
\end{figure}

A sensing strategy determines the order in which the channels are sensed and the number of users which sense a channel. In addition, the sensing strategy should also provide the timing schedule for each user to sense different channels. The optimal sensing strategy, which is called~\emph{general strategy} in this paper, includes any possible strategy to sense a set of channels. For instance, consider the scenario in Fig.~\ref{Fig:StrategiesExample}(b) with $4$ channels and $3$ users. Channel $1$, $2$, and $4$ are sensed respectively by users $1$, $2$, and $3$ starting from the beginning of the slot. To sense channel $3$, there are $3$ possibilities: i) User $1$ solely senses channel $3$ when it finishes its job sooner than user $2$ and $3$. Hence sensing channel $3$ is finished at $T_I^{(3)}=\tau_{1,1}+\tau_{3,1}$; ii) User $1$ waits for user $3$ to finish its job and then they cooperatively sense channel $3$ and $T_I^{(3)}=\tau_{4,1}+\tau_{3,2}$; iii) Both users $1$ and $3$ wait also for user $2$ and then sense channel $3$ cooperatively and $T_I^{(3)}=\tau_{2,1}+\tau_{3,3}$. As shown in the figure, it is assumed that option (ii) is the optimal solution. However, due to the large number of possible solutions, solving for the general strategy is highly cumbersome and can not be done efficiently in a timely manner, and it is also difficult to be implemented in practice. Therefore, in this section, to simplify the general sensing strategy, we propose three classes of multi-channel sensing strategies with particular structures. Each strategy can be considered as a sub-optimal scheme for the general strategy. In other words, we assume specific sensing strategies and, given this strategy, we provide the optimal answer. We assume in this section that we have homogeneous sensors with the same detected SNR~$\gamma$. 

The three proposed strategies are: i) a~\emph{sequential strategy} where all channels, which can be sensed in $T$, are sensed cooperatively by all $N$ users in a sequential manner, ii) a~\emph{parallel strategy} where channels are cooperatively sensed in parallel with a subset of users, and iii) a mixture of sequential and parallel strategies  called~\emph{sequential-parallel strategy} where different sets of channels are sensed in parallel, but channels in each set are sensed in a sequential manner. An example for each strategy is provided in Fig.~\ref{Fig:StrategiesExample}. 
\subsection{Sequential Sensing Strategy}
\label{sec:sequential}
The sequential strategy with cooperative sensing was first discussed in~\cite{liu13}. We briefly review this strategy and show an example in Fig.~\ref{Fig:StrategiesExample}(d). Given a list of users and channels, it is assumed that all users cooperatively sense each channel, and channels are thus sensed one by one. Therefore, the cooperative sensing time of any channel $m$ by $N$ sensors is given by $\tau_{m,N}$. The sensing order is defined as $\mathcal{A}=(a_1, a_2, \dots, a_M)$ which is a permutation of $(1,2,\dots,M)$, e.g., $a_1=3$ implies that the first channel being sensed cooperatively by all users is channel $3$. The expected spectrum opportunity throughput can then be re-written from equation~(\ref{eq:Throughput}) as
\begin{align}
\max\limits_{\mathcal{A}} {\mathbb{E}\{R(\mathcal{A})\}=\sum_{i=1}^{M}{\frac{(T-T_I^{(a_i)})C_{a_i}(1-u_{a_i})}{T}}},
\label{eq:throughput_sequential}
\end{align}
where $T_I^{(a_i)}=\min\left\{\sum_{j=1}^{i}{\tau_{a_j,N}},T\right\}$. In~\cite{liu13}, it is proved that the optimal sensing order is found by sorting the channels in decreasing order of $\frac{C_j(1-u_j)}{\tau_{j,N}}, j=1,\dots,M$. 
\subsection{Parallel Sensing Strategy}
\label{sec:parallel}
In this strategy, channels are sensed in parallel and the central controller makes the decision on the number of users who should sense each channel. Intuitively, when no user is assigned to a channel, the channel is not sensed and no spectrum opportunity throughput is available for this channel. Each user thus senses only one channel. An example of this strategy is illustrated in Fig.~\ref{Fig:StrategiesExample}(c). The optimization problem can be represented as
\begin{align}
&\max\limits_{\mathcal{A}}{\mathbb{E}\{R(\mathcal{A})\}=\sum_{i=1}^{M}{\frac{(T-\tau_{i,k_i})C_{i}(1-u_{i})}{T}},}~\mbox{s.t.}~ \sum_{i=1}^{M}{k_i}=N, 
\label{eq:parallel-objective}
\end{align}
where $k_i$ is the number of users assigned to channel $i$ and $\mathcal{A}=\mathbf{k}=(k_1,k_2,\dots,k_M)$. This is a classical integer programming problem. In the following, we first discuss a dynamic programming (DP) solution, and then a heuristic solution. At the end, the condition to have an integer assignment is relaxed and a relaxed optimization problem is discussed.  
\subsubsection{Dynamic Programming}
\label{sec:DP}
As a resource allocation problem, we propose the following dynamic programming (DP) solution to find the optimal assignment~\cite{bertsekas05}. The~\emph{stage} of the DP is the channel number. Thus, starting from channel $1$, we must decide at each stage, how many of the remaining users should be assigned to the particular channel considered. The decision variable is the number of users, the instantaneous payoff is the throughput which may be obtained from this channel, and the value function $v_k(n)$ is the total expected throughput which can be obtained from the optimal assignment from now on when $k$ channels and $n$ users remain. Transition possibilities naturally depend on the remaining number of users. Then, the Bellman equation can be written as
\begin{equation}
v_k(n)=\max\limits_{0\leq j\leq n}{\Bigg{\{} \frac{(T-\tau_{k,j})C_k(1-u_k)}{T}+v_{k-1}(n-j) \Bigg{\}}}.
\label{eq:DP}
\end{equation}
The terminal condition is when no users remain to be assigned, i.e., $v_k(0)=0$, $\forall k$. We thus have $v_{1}(n)=\frac{(T-\tau_{M,n})C_M(1-u_M)}{T}$, meaning that in the last stage, any remaining users should be assigned to the last channel (channel $M$). The DP is finite, so it is solved by backward induction, and the maximal throughput is equal to $v_M(N)$. Note that since the channels are sensed in parallel, sensing order and the order of channels in the DP are irrelevant.

In the proposed DP solution, we choose the users as resource to be assigned since the DP has a lower runtime complexity compared to the case where channels are assigned. Consider an optimal assignment $\mathcal{A}^{o}=(k_1,\dots,k_M)$ with a given number of users. Assume one new user is assigned to channel $i$ to achieve an optimal allocation; for any other channel $j$, we should thus have the condition $\mathbb{E}\{R(k_1,\dots,k_i+1,\dots,k_M)\}\geq\mathbb{E}\{R(k_1,\dots,k_j+1,\dots,k_M)\}$, $\forall j\neq i$, and it can be simplified as $(\tau_{i,k_i}-\tau_{i,k_{i}+1})C_i(1-u_i) \geq (\tau_{i,k_j}-\tau_{i,k_j+1})C_j(1-u_j)$. Therefore, each new user is added to a channel $i$ with currently $k_i$ assigned users which has the highest $(\tau_{i,k_i}-\tau_{i,k_i+1})C_i(1-u_i)$ value. The DP  algorithm for the parallel sensing strategy is presented in Algorithm~\ref{alg-parallel}.
\begin{algorithm}
\footnotesize
\caption{Pseudo Algorithm for Parallel DP Solution}
\label{alg-parallel}
\begin{algorithmic}[1]
\FOR {$m=1:M$}
\FOR {$n=1:N$}
\STATE $R(m,n)=(T-\tau_{m,n})C_{m}(1-u_{m})$
\ENDFOR
\ENDFOR
\STATE $R(:,0)=0$
\STATE $\mathcal{A}=0$
\WHILE {$N>0$} 
\STATE $m^*= \text{arg}\max\limits_{m} \Delta{R}=R(m,k_m+1)-R(m,k_m)$, $1\leq m\leq M$
\STATE $k_{m^*}=k_{m^*}+1$
\STATE $N=N-1$
\ENDWHILE
\end{algorithmic}
\normalsize
\end{algorithm}
\normalsize
\subsubsection{Greedy Heuristic}
\label{sec:GH}
The high execution complexity of the DP solution prompts the need to have a low-complexity heuristic. A simple, yet efficient solution is a greedy heuristic that puts more users on a channel with a higher product of the channel capacity $C_i$ and the probability of availability $1-u_i$. We thus propose
\begin{align}
k_i=\left[N \frac{C_i(1-u_i)}{\sum_{j=1}^{M}{C_j(1-u_j)}}\right],
\label{eq:heurist-parallel}
\end{align}
where $[\cdot]$ is the rounding operation. Since the sum of $k_i$ values derived from equation~(\ref{eq:heurist-parallel}) is not necessarily $N$, if $N-\sum_{j=1}^{M}{k_j}>0$ the remaining $N-\sum_{j=1}^{M}{k_j}$ users are assigned to the channel with the maximum $C_j(1-u_j)$, otherwise $\sum_{j=1}^{M}{k_j}-N$ additional users are eliminated from the channels starting with maximum $C_j(1-u_j)$. We refer to this heuristic in the figures as ``Par-GH (An.)" and ``Par-GH (Sim.)". 
\subsubsection{Constraint Relaxation}
\label{sec:relax}
In this section, we propose to relax the constraint of the optimization problem in equation~(\ref{eq:parallel-objective}) where $k_i$ is not necessary an integer. This helps us to derive a bound for the parallel strategy. It can be easily shown that the objective function $\mathbb{E}\{R(\mathcal{A})\}$ is not a simple concave function, yielding to a non-convex optimization programming solution that, given the reduced size, it can still be optimally solved by brute-force search. The detailed derivation for relaxed optimal $k_i$ are provided in Appendix~\ref{apx-relaxation}.
\subsection{Analytical Comparisons for Sequential and Parallel Strategies with Homogeneous Channels}
\label{subsec:analytical-discuss}
In this section, we compare the analytical throughput performance for the sequential and parallel strategies assuming that all channels have the same capacity $C$ and channel occupancy rate $u$, i.e., the channels are homogeneous. It is complex to analytically derive the throughput performance for heterogeneous channels since it depends on multiple channel capacities and occupancy rates. We therefore only focus our effort on obtaining analytical results for the homogeneous case to gain a better insight on the conditions, e.g., number of channels, number of users, capacity, and occupancy rate, which make one scheme better than the other, as fewer variables are involved. Let us start with the parallel strategy. 

In the parallel scheme, it can be observed that in practical scenarios, it is always better to sense more channels than to cooperatively sense fewer channels. For the case of similar channels, assume there are two channels and two users ($M=N=2$). The throughput when each user senses a channel (no sensing cooperation) can be given by $2C(1-u)(T-\tau_1)$ (channel index was dropped). The throughput of cooperatively sensing only one channel is given by $C(1-u)(T-\tau_2)$. Cooperatively sensing the same channel in the parallel strategy is thus optimal when $C(1-u)(T-\tau_2)>2C(1-u)(T-\tau_1) \Rightarrow  2\tau_1-\tau_2 >T$. Note this condition is rarely met, so it can be claimed that when channels are similar, it is better, for the parallel scheme, to distribute the users as much as possible to sense and exploit more channels. Given this insight and based on what we observed in Fig.~\ref{GainVsUserIncrease.pdf}, the total throughput of the parallel scheme is represented as
\begin{equation}
\mathbb{E}\{R_{\text{Hom}}^{\text{Par}}\}= \frac{(M-r)[C(1-u)(T-\tau_{L})] + r[C(1-u)(T-\tau_{L+1})]}{T},
\label{eq:par-hom-anal}
\end{equation}
where $L=\lfloor \frac{N}{M} \rfloor$, $\lfloor\cdot\rfloor$ is the floor operator, and $r$ is the reminder of the division, i.e., $r=\text{mod}(N,M)$. The intuition is as follows. We should first assign $L$ users to each channel and the remaining $r$ users are distributed among $r$ channels, so $r$ channels will be sensed by $L+1$ users and the others by $L$ users. We can also derive the analytical throughput performance of the parallel sensing greedy heuristic algorithm for homogeneous channels. Since channels are assumed similar, users are evenly assigned to channels and the remaining users are assigned to one of the channels. Defining $Q=[\frac{N}{M}]$, the throughput obtained by this heuristic can thus be given by:
\begin{align}
&\mathbb{E}\{R_{Hom}^{Par-GH}\}=\notag\\
&\!\!\!\!\!\!\begin{cases}
\frac{N(T-\tau_1)C(1-u)}{T} &\text{if}~N<M,\\
\frac{M(T-\tau_Q)C(1-u)}{T} &\text{if}~MQ=N~\text{and}~N\geq M,\\
\frac{(M-1)(T-\tau_Q)C(1-u)}{T}\\
+\frac{(T-\tau_{N-(M-1)Q})C(1-u)}{T} &\text{if}~MQ<N~\text{and}~N\geq M,\\
\frac{(M-t)(T-\tau_Q)C(1-u)}{T}\\
+\frac{(T-\tau_{Q-MQ+N-(t-1)Q})C(1-u)}{T} &\text{if}~MQ>N, N+(t-1)Q<MQ,\\
&MQ\leq N+tQ,~\text{where}~t\in\mathbb{Z}^+,\\
&\text{and}~N\geq M.
\label{eq:Par-GH_An}
\end{cases}
\end{align}

For the sequential model, the sensing time of each channel is $\tau_{N}$, so at most $\lfloor \frac{T}{\tau_N} \rfloor$ channels can be sensed. Let us define $K=\min\{\lfloor \frac{T}{\tau_N} \rfloor,M\}$. The total throughput is thus given by
\begin{align}
\mathbb{E}\{R_{\text{Hom}}^{\text{Seq}}\}=\frac{C(1-u)\sum\limits_{i=1}^K(T-i\tau_N)}{T}=\frac{KC(1-u)\left(T-\frac{K+1}{2}\tau_N\right)}{T}
\label{eq:hom-seq-anal}
\end{align}
Using equations~(\ref{eq:par-hom-anal}) and~(\ref{eq:hom-seq-anal}), we are able to find the operating regions where one of the strategies outperforms the other, as will be illustrated in Section \ref{sec:numerical_results}. 
\subsection{Sequential-Parallel Strategy}
\label{sec:SP}
We propose in this section a hybrid strategy named sequential-parallel. As can be seen in the example provided in Fig.~\ref{Fig:StrategiesExample}(e), channels are divided into several subsets, where within each channel subset, a subset of users are adopting sequential cooperative sensing. In other words, within each channel subset, a sequential strategy is followed while different channel subsets are sensed in parallel. The decision to be made is thus to find the channel subsets, the assignment of users to each subset and the sequential sensing order within each subset.

We define a function $R_s(\mathcal{S}_m,n)$ which is the maximum expected throughput obtainable from sequentially and cooperatively sensing by $n$ users the channel subset $\mathcal{S}_m$. From Section~\ref{sec:sequential}, we already have the optimal sequential strategy within one channel subset. With this type of structure, the throughput maximization problem is indeed a Knapsack problem \cite{martello90} where we are looking for the best 2-tuples $(\mathcal{S}_m,n)$ to put in the knapsack. In the following, a dynamic programming model and a greedy heuristic are proposed to solve this problem.   
\subsubsection{Dynamic Programming}
\label{sec:DP_SP}
Given the function $R_s$, the state variable in the DP equation is represented by $(\mathcal{S},n)$, where $\mathcal{S}$ is a subset of channels, not sensed yet, and $n$ is the number of remaining users, not assigned to any channel set. The decision is one of the subsets of $\mathcal{S}$ and the number of users assigned to it. Therefore, the total number of possible actions is equal to $2^{|\mathcal{S}|}(n+1)$. The Bellman equation can be given by
\begin{equation}
v(\mathcal{S},n)=\max\limits_{0\leq j\leq n, \mathcal{X} \subseteq \mathcal{S}}{\Bigg{\{} R_s(\mathcal{X},j)+v(\mathcal{S}-\mathcal{X},n-j) \Bigg{\}}},
\label{eq:Seq-Par_DP}
\end{equation}
where $\mathcal{X}$ is the decision variable which is a subset of $\mathcal{S}$. The DP model is of infinite-horizon, so it can be solved by value iteration~\cite{bertsekas05}. As soon as we reach any state with $v(\emptyset,n)$ or $v(\mathcal{S},0)$, the ongoing payoff is zero and the solution is terminated. 
\subsubsection{Greedy Heuristic}
\label{sec:GH_SP}
Similar to the classical Knapsack problem, the greedy approach starts with the $2$-tuple whose ratio of throughput versus the number of users is maximum. When a channel subset and the number of users assigned to this subset are decided, the algorithm is continued for the remaining users and channels. The greedy algorithm can be found in Algorithm~\ref{alg-seq-parallel-heuristic}. As discussed in~\cite{martello90}, the greedy heuristic is guaranteed to have a performance higher than half of the optimal result.
\begin{algorithm}
\footnotesize
\caption{Pseudo Algorithm for Sequential-Parallel Greedy Heuristic Solution}
\label{alg-seq-parallel-heuristic}
\begin{algorithmic}[1]
\WHILE {$N>0$} 
\STATE Select 2-tuple $(\mathcal{S}^*,n^*)$ with maximum $\frac{R_s(\mathcal{S}_m,n)}{n}$
\STATE Remove all entries $(\mathcal{S}_m,n)$ if $\mathcal{S}_m \cap \mathcal{S}^* \neq \emptyset$
\STATE Remove all entries $(\mathcal{S}_m,n)$ if $n>N-n^*$
\STATE $N=N-n^*$
\ENDWHILE
\end{algorithmic}
\normalsize
\end{algorithm}
\normalsize
\subsection{Iterative-Parallel Solution for the General Spectrum Sensing Strategy}
\label{sec:iterative_parallel}
The solutions provided for the parallel sensing strategy raise the idea of using an iterative solution for the general model. Let us call the solution algorithm proposed for the parallel strategy~$\emph{ParallelStrategy}(\cdot)$, which could be any of the proposed solutions. In the first iteration of the iterative approach for the general model, a decision based on the parallel strategy is made to sense some channels in parallel. Those channels are removed from the list and for the remaining channels, a new decision is made based on the parallel strategy. Iterations are continued until all channels are sensed (or until the end of time slot). However, we observe that in the parallel strategy users are all synchronized, while in the iterative solution, sensing time of the channels may be different, and hence users finish their first assigned job in different time instants. 
\begin{figure}
\centering
\includegraphics[width=0.68\columnwidth]{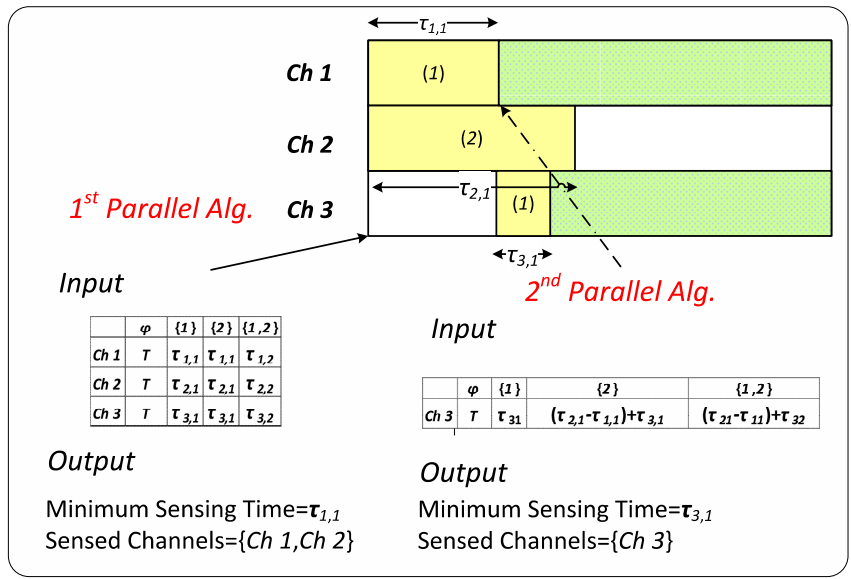}
\caption{An example of the proposed Iterative-Parallel solution for the general sensing model.}
\label{IterativeParallel.pdf}
\end{figure}

To be able to employ the parallel strategy iteratively, we use the following approach, which is illustrated with an example in Fig.~\ref{IterativeParallel.pdf}. The parallel strategy has been called in the beginning of the time slot and the optimal decision is to sense channel $1$ by user one and channel $2$ by user two. Now, consider the point that sensing the first channel (shortest sensing time) is finished and user one becomes idle. We call this point the new reference point where a new decision is made. The remaining sensing time of channel $2$ is known, so if a new job (sensing channel $3$) is assigned to a designated user who is currently busy, we have to wait first for this user to finish its current job which takes $(\tau_{21}-\tau_{11})$ time unit, and then the user starts the new job. It is equivalent to assume that the sensing time of any remaining channel, here channel $3$, by this designated user, starting at new reference point, is the sum of remaining sensing time of the channel being sensed by this designated user and the original sensing time of the remaining channel, which is equal to $(\tau_{21}-\tau_{11})+\tau_{31}$. Similarly, the cooperative sensing time of channel $3$ is updated to $(\tau_{21}-\tau_{11})+\tau_{32}$ because user one should wait for user two to finish and then join for a cooperative sensing.  
For cooperative sensing, as discussed in Section~\ref{sec:cooperative_sensing}, all collaborators should start at the same time, so the updated sensing time for a channel is defined based on the longest remaining job. As described in Algorithm~\ref{alg-iterative-parallel}, in each iteration, the list of remaining channels is updated and based on the remaining job of the users, the table of all sensing times by different subsets of users is recalculated. By modifying the length of the time slot in each iteration (as throughput function is linear versus the time slot), the reference point is redefined. Note that $\tau_s^{[]}$ is a matrix with $M\times2^N$ entries which keeps the sensing time of each channel sensed cooperatively by a subset of users, as discussed in Section \ref{sec:fusion}.  After each iteration, some channels remain which are still not sensed. Therefore, a new decision is made only considering the remaining channels. It is worth noting that this algorithm is run offline in the beginning of the time slot (similar to a DP) to find the optimal strategy, then the strategy is followed and applied to the time slot. It is clear that since we are running the parallel algorithm, we are maximizing the instantaneous payoff, so the proposed solution is a myopic solution and not necessarily optimal. 
\begin{algorithm}
\footnotesize
\caption{Pseudo Algorithm for Iterative-Parallel Solution}
\label{alg-iterative-parallel}
\begin{algorithmic}
\STATE Function IterativeParallel($N,M,\tau_s^{[]},T$).
\STATE $\mathbf{CH}=1:M$
\WHILE {IsNotEmpty($\mathbf{CH}) \hspace{1mm} \& \hspace{1mm} T>0$} 
\STATE SensingSchedule=ParallelStrategy($N$,$\mathbf{CH}$,$\tau_s^{[]}$,$T$).
\STATE $\mathbf{CH}=\mathbf{CH}$-SensedChans (SensingSchedule).
\STATE $T=T$-MinSensingTime (SensingSchedule).
\STATE $\tau^M$=UPDATE($T$,$\tau_s^{[]}$,SensingSchedule).
\STATE Solution=[Solution SensingSchedule].
\ENDWHILE
\RETURN Solution.
\end{algorithmic}
\normalsize
\end{algorithm}
\normalsize
\subsection{Memory Usage and Computational Complexity Discussion}
\label{sec:complexity}
It is not possible to solve the throughput optimization problem in the general sensing strategy in polynomial time, since all permutations of $M$ channels along with all ways to divide $N$ users among $M$ channels need to be considered. In terms of the memory space complexity, the maximum memory space required for it is $\mathcal{O}(2^M2^N)$ to keep the sensing time of any subset of channels by any subset of users, where $\mathcal{O}(\cdot)$ is the big O notation. For the other proposed strategies, the memory space and computational complexity are summarized in Table~\ref{tab:complexity}. Note that for the sequential-parallel strategy, even though their big O complexity is the same, DP has more lower order computation terms than the heuristic (e.g., $3n^3+2n^2+n$ versus $n^3$ while both are $\mathcal{O}(n^3)$). In addition, in calculating the computational complexity, the execution time to fill the required data structures, already taken into account in the memory usage, is not considered to avoid repetition.  
%

\scriptsize
\begin{table}[]
\caption{{Memory space and computational complexity of homogeneous sensing strategies.}}
\label{tab:complexity}
\centering
\begin{tabular} {c|l|l}
\hline
\scriptsize
{\textbf{Strategy}} & {\textbf {Memory}} & {\textbf {Computation}}\\
 \hline
Sequential& $\mathcal{O}(M)$ & $\mathcal{O}(M^2)$\\
 \hline
Parallel-DP& $\mathcal{O}(MN)$ & $\mathcal{O}(MN^2)/\mathcal{O}(M^2N)$\\
 \hline
Parallel-Heuristic& $\mathcal{O}(1)$ & $\mathcal{O}(M)$ \\
 \hline
Sequential-Parallel-DP& $\mathcal{O}(2^MN)$ & $\mathcal{O}(2^{2M}N^2)$\\
 \hline
Sequential-Parallel-Heuristic& $\mathcal{O}(2^MN)$ & $\mathcal{O}(2^{2M}N^2)$\\
\hline
Iterative-Parallel-DP& $\mathcal{O}(MN)$ & $\mathcal{O}(M^2N^2)$\\
\hline
\end{tabular}
\normalsize
\end{table}
\normalsize
\section{Spectrum Sensing Strategies with Heterogeneous Sensors}
\label{sec:hetsensors}
In the previous section, it was assumed that the users are homogeneous in sensing with the same detection SNR~$\gamma$ for all channels and all users. Due to different distances between the SUs and PU transmitter, as well as channel variations, sensors in practice may have different detection SNRs and consequently different sensing times to fulfill a given sensing accuracy. In this section, the optimal sensing strategy is therefore investigated assuming different detection SNR values~$\gamma_{m,i}$ for SU $i$ in channel $m$.   
\begin{figure}%
\centering
\includegraphics[width=0.68\columnwidth]{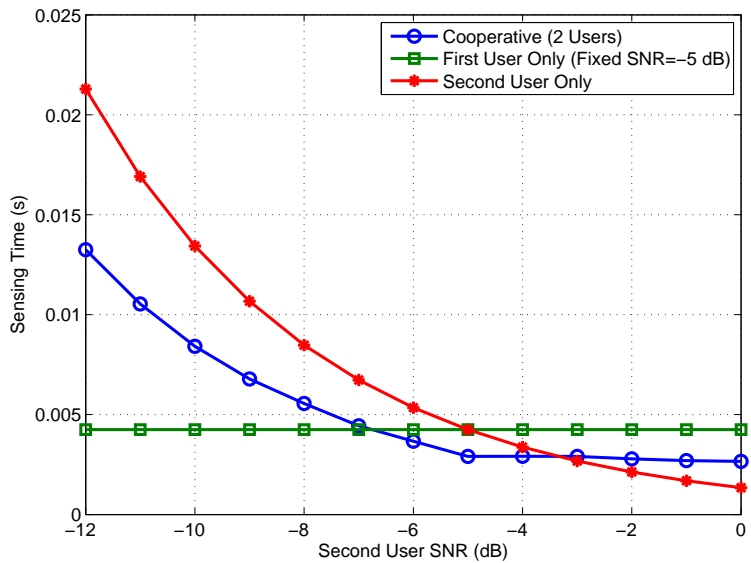}%
\caption{Cooperative sensing time by \textit{AND} fusion rule versus single-user sensing to satisfy aggregated sensing accuracy of $Q^*_d=0.9$ and $Q^*_f=0.15$. Other parameters are $f_s=4$ kHz, noise power $\sigma^2=10^{-5}$ W and the detection SNR for the first user $\gamma_1=-5$ dB.}%
\label{HetSensorsSampleFigAND2}%
\end{figure}
In this paper we assume a more realistic scenario where the cooperative sensing time is the same for all participating users and sensing reports are sent to the controller at the same time. We do not address the scenarios where the sensing duration is different for different users and the fusion center has to wait to receive the last report to make its decision. Considering this assumption, a key observation is that, depending on the fusion rule, the cooperative sensing time to meet target $Q^*_f$ and $Q^*_d$ values may even increase with additional users if a user with a significantly lower detection SNR cooperates. Fig.~\ref{HetSensorsSampleFigAND2} illustrates this fact. Only between the crossover points at $-7$ dB and $-3$ dB should the users cooperatively sense the channel to achieve the lowest sensing time. Below a $-7$ dB SNR for the second user, the channel should be sensed only by the first user, which has a significantly better SNR than the second user. Similarly, when the second user SNR is above $-3$ dB, only the second user should sense the channel for the best performance. 

For each channel $m$ and within a set of sensors $\mathcal{S}$, with known SNRs, we can thus find the optimal subset of users $\mathcal{S}^*_m \subseteq \mathcal{S}$ to cooperatively sense this channel. The cooperative sensing time for a channel $m$ will be the minimum sensing time such that the target sensing accuracy is satisfied. Let us call it $\tau^{*}_{m}$. In the following, we discuss how $\tau^{*}_{m}$ is found for a channel $m$. It should be noted that finding $\tau^{*}_{m}$ is just an initial step in the scheduling optimization procedure and we then come back in Section~\ref{sec:hetsensors-seq} to the original optimization problem, which is to maximize the potential throughput by providing solutions to our proposed scheduling strategies. 

Consider a subset $\mathcal{S}$ of all $N$ users with an SNR vector $(\gamma_{m,1}  \gamma_{m,2} \dots \gamma_{m,N})$. If we assume that all users in $\mathcal{S}$ cooperatively sense channel $m$, the minimum cooperative sensing time $\tau_{m,\mathcal{S}}$ can be found by solving the following optimization model
\begin{align}
\min\limits_{\tau,\Upsilon} \tau,~\mbox{s.t.}~Q_d \geq Q^*_d,~Q_f \leq Q^*_f,
\label{Eq:HetSensors-Opt}
\end{align}
where $\tau$ is the cooperative sensing time and $\Upsilon=(\epsilon_1 \dots \epsilon_{N})$ is the detection threshold of all users. For the matched filter detection, we have~\cite{Brodersen06} 
$P_{f,i}=Q(\frac{\epsilon_{i}}{\sqrt{\tau f_s \theta_i \sigma_w^2}})$ 
and 
$P_{d,i}=Q(\frac{\epsilon_{i}-\tau f_s \theta_i}{\sqrt{\tau f_s \theta_i \sigma_w^2}})$, 
where $\epsilon_{i}$ is the detection threshold, $\tau$ is the sensing time, $\theta_i$ is the PU signal power so that $\gamma_{i}=\frac{\theta_i}{\sigma_w^2}$,  $\sigma_w^2$ is the noise power and $f_s$ is the sampling frequency. Channel index has been dropped in all parameters above. 

For~\emph{AND} fusion rule, we also have that ${{Q_f}}=\prod_{i \in \mathcal{S}}(P_{f,i})$ and ${{Q_d}}=\prod_{i \in \mathcal{S}}(P_{d,i})$, where $P_{f,i}$ and $P_{d,i}$, $1\leq i\leq N$ are the probability of false alarm and detection for the $i$-th user in $\mathcal{S}$, respectively. Note they are both functions of $\tau$ and detection thresholds $\Upsilon$. We thus have $|S|$ equations and $|S|+1$ variables ($\tau$ and $\epsilon_{i}$), and we are looking for the minimum $\tau$ and a feasible $\Upsilon$. The optimization problem above can be solved by a solver to find the minimum cooperative sensing time $\tau_{m,\mathcal{S}}$. 

The sensing time found above is the cooperative sensing time when all users in a subset $S$ cooperate. However, as we previously discussed, increasing the number of cooperating sensors does not necessarily improve the sensing time, and the optimal set of users can be a subset of the users. Therefore, the optimal sensing time for channel $m$ can be given by $\tau^{*}_{m}= \min\limits_{\mathcal{S}\subseteq\{1,2,\cdots,N\}}{\tau_{m,\mathcal{S}}}$, where $\mathcal{S}$ is any subset of $N$ users, and the cooperative sensing time for this subset is given by equation~(\ref{Eq:HetSensors-Opt}). The complexity of finding the best subset can be decreased from $O(2^N)$ to $O(N^2)$ by the fact that it is not required to check all subsets. The user with the best SNR should necessarily be one of the cooperating sensors, so if we sort the vector of SNRs in a descending order and rename the users accordingly, only the sensing time of $N$ subsets $\{1\}, \{1,2\}, \dots \{1,2,\dots,N\}$ should be found and compared. Starting from $\{1\}$ towards larger subsets, we continue the search until the cooperative sensing time increases by adding the next user. We can see that this is equivalent to finding an SNR threshold and only have users who have an SNR greater than this threshold participate to the cooperative sensing. 

Since the complexity of solving equation~(\ref{Eq:HetSensors-Opt}) depends on the solver and the algorithm used, we call it $E(N)$. From now on, we assume that the we know the optimal sensing time and the optimal sensing set for any channel, found with $E(N)+O(MN^2)$ complexity, and the cooperative sensing time for any subset of users found with an $E(N)+O(M2^N)$ complexity. Using those pieces of information, we now discuss how the sequential, parallel and sequential-parallel sensing strategies can be designed.

\subsection{Sequential Sensing with Heterogeneous Sensors}
\label{sec:hetsensors-seq}
As discussed, for any channel $m$, we could find the optimal set of users and the optimal sensing time $\tau^{*}_{m}$. We use this optimal sensing time to find the optimal channel sensing order by sorting the channels in the decreasing value order of $\frac{C_m(1-u_m)}{\tau^*_m}$, and each channel is cooperatively sensed by its optimal set of users. Therefore, the only change compared to the case with homogeneous sensors is that a channel is not sensed by all users and a user will not thus necessarily sense all channels. The results will be called ``Seq (Opt.)" in the numerical results figures.

As solving equation~(\ref{Eq:HetSensors-Opt}) and finding the optimal sensing time may be cumbersome for a large number of users, we propose the following average-based heuristic called ``Seq-Heuristic-Avg" in the figures. For each channel $m$, we assume that the channel is sensed only by the users who have a detection SNR greater than the average SNR of all users. The rationale behind this heuristic is that, for a large number of users, there will also be a large number of users in the sensing set. As discussed previously, the cooperative sensing time gain is significant mostly for the first users. Therefore, discarding a few users will not have a major impact on the performance. We then assume that all users who have a SNR higher than the average have the same average SNR and we find the cooperative sensing time from equation~(\ref{eq:sensing_time}) denoted by $\tau^{\text{Avg}}_m$. Channels are sensed in descending order of $\frac{C_m(1-u_m)}{\tau^{\text{Avg}}_m}$. 

\subsection{Parallel Sensing with Heterogeneous Sensors}
\label{sec:hetsensors-par}
We saw that each channel has an optimal set of users to cooperatively sense that channel. If those sets were disjoint, it would be optimal to sense all channels in parallel, each one with its optimal set. Since the sets are not necessarily disjoint, we propose the following assignment strategies which follow similar concepts to the ones proposed for homogeneous sensors. Using the cooperative sensing time for any channel and by any set of users given from equation~(\ref{Eq:HetSensors-Opt}), we could have a dynamic programming model to find the optimal assignment, and a sub-optimal scheme with a lower complexity similar to the one proposed in Algorithm~\ref{alg-parallel}. Note that since the sensors are heterogeneous, the latter approach is not the solution of the DP (as it was for homogeneous sensors) and results in a sub-optimal assignment (i.e., the order of user assignment to channels is important here). We call the optimal and heuristic approaches respectively ``Par-DP (Opt.)" and ``Par-DPSim-HetSens" in the numerical results figures. 

We also propose a classical greedy scheme, using the average sensing time introduced for sequential strategy. The channels are sorted in decreasing order of $\frac{C_m(1-u_m)}{\tau^{\text{Avg}}_m}$. The best channel is selected and the user with the best SNR for this channel is assigned to the channel and removed from the list of users. The algorithm is then continued for other channels. If one user is assigned to all channels and there are still unassigned users (i.e., $M<N$), the assignment is restarted from the first channel to add another sensor. This approach is called ``Par-Avg-Greedy-HetSens" in the numerical results. As can be seen, the objective of this heuristic is to explore the most channels in parallel with their best user sets.

\subsection{Sequential-Parallel Sensing with Heterogeneous Sensors}
\label{sec:hetsensors-seqpar}
Similar to the Section~\ref{sec:SP} with homogeneous channels, we can find the expected throughput which can be obtained from a set of channels sensed cooperatively by a set of users in a selective manner. This is equivalent to using the sequential approach proposed above for any set of users and channels. Recall that due to heterogeneity in sensing, all users may not sense all channels. Forming a matrix with an $O(2^M 2^N)$ complexity, we could have a DP model to find the optimal assignment. It is called ``Seq-Par-DP (Opt.)" in the numerical results figures. The complexity will however be too high. In the following, we thus propose a heuristic with a very low complexity. 

In this heuristic, called ``Seq-Par-Heuristic" in the numerical results figures, channels are sorted based on the obtainable throughput divided by the number of sensors, given the optimal sensing time. The first channel which is selected thus has the maximum $\frac{R_m(\tau^*_m)}{|\mathcal{S}^*_m|}$ where $R_m(\tau^*_m)$ is the throughput obtainable from channel $m$ if it is sensed by its optimal set of users. After selecting the first channel, any other channel $i$ whose optimal set of sensors is a subset of the optimal set of the selected channel (i.e., $\mathcal{S}^{*}_i \subseteq \mathcal{S}^*_m$) is also selected to be sensed by the same set of sensors in a sequential manner. All these channels are removed from the list and a new decision is made for the remaining channels and users. Naturally, the optimal set of users and optimal sensing time for all remaining channels must then be recalculated based on the remaining users. 

\subsection{Memory Usage and Computational Complexity Discussion with Heterogeneous Sensors}
\label{sec:complexity-het}
The complexity comparison of different solutions for the heterogeneous cases is summarized in Table~\ref{tab:complexity-het}. The main difference between homogeneous and heterogeneous cases from complexity point of view is the extra calculation for equation~(\ref{Eq:HetSensors-Opt}) for each channel, which is denoted by $E(N)$. In other words, after finding the optimal sensing time and optimal sets for each channel, the complexity is similar to homogeneous scenarios.  

\scriptsize
\begin{table}[]
\caption{{Memory space and computational complexity of heterogeneous sensing strategies.}}
\label{tab:complexity-het}
\centering
\begin{tabular} {c|l|l}
\hline
\scriptsize
{\textbf{Strategy}} & {\textbf {Memory}} & {\textbf {Computation}}\\
 \hline
Sequential (Opt)& $\mathcal{O}(M N E(N))$ & $\mathcal{O}(M^2)$\\
 \hline
Seq-Heuristic-Avg& $\mathcal{O}(M)$ & $\mathcal{O}(M^2)$\\
 \hline
Parallel (DP,Opt)& $\mathcal{O}(M 2^N E(N))$ & $\mathcal{O}(M 2^{2N})$ \\
 \hline
Parallel-DPSim-Het& $\mathcal{O}(M 2^N E(N))$ & $\mathcal{O}(M N^2)$\\
 \hline
Parallel-Avg-Greedy& $\mathcal{O}(MN)$ & $\mathcal{O}(M^2 N)$\\
\hline
Sequential-Parallel (DP,Opt)& $\mathcal{O}(2^M 2^N E(N))$ & $\mathcal{O}(2^M 2^{2N})$\\
\hline
Sequential-Parallel-Heuristic& $-$ & $\mathcal{O}(M^2 N E(N))$\\
\hline
\end{tabular}
\normalsize
\end{table}
\normalsize

\section{Robust Spectrum Sensing Scheduling Design}
\label{sec:robust_scheduling}
Observing equation~(\ref{eq:Throughput}), it can be seen that three system parameters, i.e., PUs duty cycle $u$, detection SNR $\gamma$, embedded in sensing time, and received SNR of SU transmission, embedded in capacity, have a random nature while in previous sections, we assumed a perfect knowledge of those parameters. A robust optimization, maximizing the throughput while keeping its variation below a threshold, can be provided for those parameters to take into account their random nature. Since in this paper we focused on the sensing without taking the actual transmission into account, we continue to use the assumption of full knowledge of the channel capacity and discuss how the variation of the two other random parameters affect the decision made by different sensing strategies. 

We now separately relax the assumptions that we have full knowledge for the duty cycle $u$ and the detected SNR~$\gamma$, i.e., we first analyze the case with imperfectly known duty cycle while the detected SNR is still assumed perfectly known, and then the inverse case is investigated. In the actual system, we need to estimate those parameters accurately so that the estimates would not degrade the system performance tremendously. In the following robust system design, we first propose a low-complexity average estimator to estimate the duty cycle $u$ on a single channel, based on the discrete-time Markov chain (DTMC) assumption for PU transmission traffic~\cite{liu13}, and then we analyze the statistics for the proposed estimator. Second, we formulate two uncertainties, i.e., primary traffic and channel uncertainty, and combine them with our proposed cooperative spectrum sensing scheduling schemes as a joint robust optimization problem. The primary traffic uncertainty comes from the estimation errors of the duty cycle, and the channel uncertainty comes from the detected SNR, which is not a constant but a random variable following a certain distribution.
\subsection{Primary Traffic Estimation: Average Estimator and Its Performance Analysis}
\label{sec:traffic_estimation}
The PU traffic on a single channel is modeled as a DTMC~\cite{Cabric13}, where $z$ represents the current state of the PU (also termed as a PU traffic sample). States $z=0$ and $z=1$ indicate the PU is absent and present, respectively within one slot time $T$. This traffic is characterized by the steady-state distribution and the transition probabilities. The probability of PU absence is denoted as $P_0=\Pr\{z=0\}=1-u$ and of PU presence as $P_1=\Pr\{z=1\}=u$. The transition probabilities from state $x$ to state $y$ is denoted as $P_{xy}$ with four probabilities $\{P_{00},P_{01},P_{10},P_{11}\}$ with $u=\frac{P_{01}}{P_{01}+P_{10}}$, $P_{00}+P_{01}=1$, and $P_{10}+P_{11}=1$. 

Assume we obtain $W$ PU traffic samples to form a vector $\mathbf{z}=(z_1,z_2,\cdots,z_W)$, $z_i\in\{0,1\}$, $1\leq i\leq W$ under perfect sensing from a PU channel. Our goal is to estimate the duty cycle $u$ using these observed traffic samples $z_i$. We adopt a low-complexity average estimator, i.e., $\hat u=\frac{1}{W}\sum\limits_{i=1}^{W}z_i$~\cite{Cabric13}. Its expected value is shown to be $\mathbb{E}\{\hat u\}=\frac{1}{W}\sum\limits_{i=1}^{W}\mathbb{E}\{z_i\}=u$, and therefore it is an unbiased estimator. To derive its variance, we apply the results in~\cite[Eq. (15)]{Cabric13} to obtain 
\begin{align}
\text{Var}\{\hat u\}=\frac{u(1-u)}{W}+\frac{2u(1-u)r(r^W-Wr+W-1)}{W^2(1-r)^2},
\label{Var_u}
\end{align}
where $r=P_{11}-P_{01}$\footnote{Note that~\cite[Eq. (15)]{Cabric13} is the variance for $u$ assuming the traffic samples follow a continuous-time Markov chain (CTMC). However, if we constraint the CTMC by using uniform sampling, it would turn into the DTMC, where we can simply replace $\Gamma_u$ in~\cite[Eq. (15)]{Cabric13} with $r$ in~(\ref{Var_u}). The detailed derivation can be found in Appendix~\ref{variance_estimator}.}. 
Note that the asymptotic value for the variance of $\hat u$ is $\lim_{W\rightarrow\infty}\text{Var}\{\hat u\}=0$, which means that as we  increase the number of PU traffic samples, the variance for the estimator will go to zero to make a perfect estimation.
\subsection{Robust Optimization}
\label{sec:robust_optimization}
In this section, we formulate the robust optimization problem regarding the primary traffic uncertainty and channel uncertainty, i.e., the estimation errors for duty cycle and statistical behavior from the detected SNR, respectively. Note that we relax one variable at a time and keep the other one fixed so that we can observe their effect separately.
\subsubsection{Primary Traffic Uncertainty}
\label{sec:traffic_uncertainty}
By using the estimates of duty cycle $\hat u_i$ in the proposed system, where the channel activity is assumed independent among channels, with given traffic samples $W_i$, $1\leq i\leq M$ for all channels, we first formulate the robust optimization problem by maximizing the expected estimated throughput, subject to the constraints that the variance for the estimated throughput should be no greater than a given threshold $\eta$~\cite{Giannakis08}, i.e., 
\begin{optimization}
\begin{align}
&\max\limits_{\mathcal{S}}{\mathbb{E}\{\hat R(\mathcal{S})\}=\sum_{i=1}^{M}{\frac{(T-T_I^{(i)}(\mathcal{S}))C_{i}(1-\mathbb{E}\{\hat u_{i}\})}{T}},} \nonumber \\
&\mbox{s.t.}~\mathcal{S}\in\mathcal{A},~
\text{Var}\{\hat R(\mathcal{S})\}=\sum_{i=1}^{M}{\frac{(T-T_I^{(i)}(\mathcal{S}))^2C_{i}^2\text{Var}\{\hat u_{i}\}}{T^2}}\leq\eta,
\label{eq:Robust_Traffic}
\end{align}
\label{optimization:traffic}
\end{optimization}
where $\hat R(\mathcal{S})$ is the estimated throughput and $\mathcal{S}$ is an element in the sensing strategy set $\mathcal{A}$. Since the estimator for the duty cycle is an unbiased estimator, i.e., $\mathbb{E}\{\hat u_i\}=u_i$, we have $\mathbb{E}\{\hat R(\mathcal{S})\}=R(\mathcal{S})$. In addition, we can obtain $\text{Var}\{\hat R(\mathcal{S})\}$ by substituting the equation~(\ref{Var_u}) into equation~(\ref{eq:Robust_Traffic}). To solve this  robust optimization problem (ROP), we search for all the possible elements $\mathcal{S}$ in $\mathcal{A}$, sort the corresponding estimated throughput in a descending order, and then adopt this order to search for the variance constraint until we have  the variance being less or equal to $\eta$. This optimal solution is summarized in Algorithm~\ref{alg-optimization1}.

From a practical system design point of view, we do not have the information of how many traffic samples should we use in advance. Therefore, we need to have another optimization problem formulation to obtain an efficient system design. As observed in ROP~\ref{optimization:traffic}, the objective function $\mathbb{E}\{\hat R(\mathcal{S})\}$ does not depend on $W_i$, which means that we can always achieve the optimal solution without considering the variance constraint. However, 
since $\text{Var}\{\hat R(\mathcal{S})\}$ is a decreasing function in terms of $W_i$, as we keep increasing $W_i$, we can achieve any arbitrary variance constraint if we have sufficient traffic samples. Hence, we propose to obtain the minimum number of traffic samples if we have the variance constraint, or to obtain the minimum variance of estimated throughput if we have a sensing energy constraint, which are listed as two ROPs as follows respectively.
\begin{optimization}
\begin{align}
\min\limits_{W_i}\sum\limits_{i=1}^{M}W_i,~\mbox{s.t.}~\text{Var}\{\hat R(\mathcal{S})\}\leq\eta,
\end{align}
\label{optimization:Variance}
\end{optimization}
and
\begin{optimization}
\begin{align}
\min\limits_{W_i}\text{Var}\{\hat R(\mathcal{S})\},~\mbox{s.t.}~\sum\limits_{i=1}^{M}W_i\leq\epsilon.
\end{align}
\label{optimization:energy}
\end{optimization}
Here we show that the above two ROPs are equivalent by the following Lemma.
\begin{lemma}
ROP~\ref{optimization:Variance} and ROP~\ref{optimization:energy} are equivalent, i.e., the optimal solution for both problems are the same.
\label{lemma:equivalence}
\end{lemma}
\begin{IEEEproof}
See Appendix~\ref{proof_of_equivalence}.
\end{IEEEproof}
From the above Lemma, we can solve either ROP~\ref{optimization:Variance} or ROP~\ref{optimization:energy} to have an efficient system design by minimizing both sensing energy and variation for the throughput.
\begin{algorithm}
\footnotesize
\caption{Pseudo Algorithm for Solving ROP~\ref{optimization:traffic}}
\label{alg-optimization1}
\begin{algorithmic}[1]
\FOR {$i=1:|\mathcal{A}|$}
\STATE $Q_i=\mathbb{E}\{\hat R(\mathcal{S}_i)\}$.
\ENDFOR
\STATE Sort $Q_i$ in a descending order with the corresponding strategy $\mathcal{E}=(\mathcal{E}_1,\mathcal{E}_2,\cdots,\mathcal{E}_{|\mathcal{A}|})$, $\mathcal{E}_i\in\mathcal{A}$.
\STATE $i=1$.
\WHILE {$\text{Var}\{\hat R(\mathcal{E}_i)\}>\eta$}
\IF {$i\geq M$}
\RETURN no solution.
\ELSE
\STATE $i=i+1$.
\ENDIF
\ENDWHILE
\RETURN optimal strategy $\mathcal{E}_i^{*}$.
\end{algorithmic}
\normalsize
\end{algorithm}
\normalsize
\subsubsection{Channel Uncertainty for PU Detection}
\label{sec:channel_uncertainty_detection}
In this section, we relax the assumption that the detected SNR $\gamma$ is a constant value for all SUs. Instead, we model $\gamma$ as a random variable for all SUs and we would like to discuss how its variation will affect the sensing scheduling optimization and thus our system performance. First, assume the channel gain $g$ from PU to SU receiver follows a Rayleigh distribution. We can easily show that $\gamma=\frac{\Psi g^2}{\sigma^2}$ follows an exponential distribution~$f(\gamma)=\beta e^{-\beta\gamma}$ if $\gamma\geq0$, else $f(\gamma)=0$ with a parameter $\beta$. Second, we make an assumption that the random variable $\gamma$ should be constrained within a certain lower bound and upper bound, i.e., $\gamma\in(\phi_L,\phi_U)$. The lower bound is due to the fact that the PU detector suffers from some channel uncertainty effects (e.g. frequency offset and noise uncertainty), hence it can not detect PU signal below a SNR value which is called the~\emph{SNR wall}~\cite{Brodersen06}. The upper bound comes from the fact that the PU detector shown in equation~(\ref{eq:sensing_time}) only targets the range of low SNR much less than $0$ dB. Hence, the truncated probability density function (PDF) for $\gamma$ can be derived as $f(\gamma)=\frac{\beta e^{-\beta\gamma}}{\int_{\phi_L}^{\phi_U}f(\gamma)d\gamma}=\frac{\beta e^{-\beta\gamma}}{e^{-\beta\phi_L}-e^{-\beta\phi_U}}$, $\gamma\in(\phi_L,\phi_U)$, else $f(\gamma)=0$.
%
%
The mean for this truncated $\gamma$ can be derived as
%
$\mathbb{E}\{\gamma\}=\frac{1}{\beta}+\frac{\phi_Le^{-\beta\phi_L}-\phi_Ue^{-\beta\phi_U}}{e^{-\beta\phi_L}-e^{-\beta\phi_U}}$.
%
Followed by the above assumptions, we can formulate a similar optimization problem as shown in ROP~\ref{optimization:traffic} which is to maximize the expected throughput given the variance constraint under the SNR variation, i.e., 
\begin{optimization}
\begin{align}
&\max\limits_{\mathcal{S}}{\mathbb{E}\{R(\mathcal{S})\}=\sum_{i=1}^{M}{\frac{\left(T-\mathbb{E}\left\{\frac{A^{(i)}(\mathcal{S})}{\gamma}\right\}\right)C_{i}(1-u_{i})}{T}},}\notag\\
&\mbox{s.t.}~\mathcal{S}\in\mathcal{A},~\text{Var}\{R(\mathcal{S})\}=\text{Var}\left\{\frac{1}{\gamma}\right\}\left[\sum_{i=1}^{M}\frac{\{A^{(i)}(\mathcal{S})\}C_{i}(1-u_{i})}{T}\right]^2\leq\eta,
\label{eq:Robust_Channel}
\end{align}
\label{optimization:SNR}
\end{optimization}
since the sensing time can be written as $T_I^{(i)}(\mathcal{S})=\frac{A^{(i)}(\mathcal{S})}{\gamma}$, where $A^{(i)}(\mathcal{S})$ is a constant value depending on the sensing strategy and the fusion rule. In order to obtain the first and second moments for the throughput, we apply the following Lemma.
\begin{lemma}
Given a random variable $\gamma$ with its truncated exponential PDF $\bar f(\gamma)$ for $\gamma\in(\phi_L,\phi_U)$ with the shape parameter $\beta$, we can obtain the first and second moments for its inverse random variable respectively as $\mathbb{E}\left\{\frac{1}{\gamma}\right\}=\frac{\beta\left(\text{Ei}(\beta\phi_L)-\text{Ei}(\beta\phi_U)\right)}{e^{-\beta\phi_L}-e^{-\beta\phi_U}}$, and $\mathbb{E}\left\{\frac{1}{\gamma^2}\right\}=\frac{\beta}{e^{-\beta\phi_L}-e^{-\beta\phi_U}}[\frac{e^{-\beta\phi_L}}{\phi_L}-\frac{e^{-\beta\phi_U}}{\phi_U}
-\beta(\text{Ei}(\beta\phi_L)-\text{Ei}(\beta\phi_U))]$, where $\text{Ei}(x)=\int_{x}^{\infty}\frac{e^{-t}}{t}dt$ is the exponential integral.
\label{Lemma:statistics_SNR}
\end{lemma}
\begin{IEEEproof}
For the first moment of $\frac{1}{\gamma}$, it can be derived by calculating it directly by the definition and with change of variable technique. For the second moment of $\frac{1}{\gamma}$, it can be derived using integral by parts combining with the result of its first moment.
\end{IEEEproof}
From Lemma~\ref{Lemma:statistics_SNR}, we can derive the variance for $\frac{1}{\gamma}$ and therefore the variance for the throughput. In addition, to solve ROP~\ref{optimization:SNR}, we can apply the same Algorithm~\ref{alg-optimization1} by simply replacing the objective and subjective functions accordingly.
\section{Numerical Results}
\label{sec:numerical_results}
In this section, numerical results evaluating the expected normalized throughput from spectrum opportunities under different strategies are presented and discussed.  
\subsection{Throughput Comparisons for All Strategies}
\begin{figure}[!t]
\centering
\includegraphics[width=0.68\columnwidth]{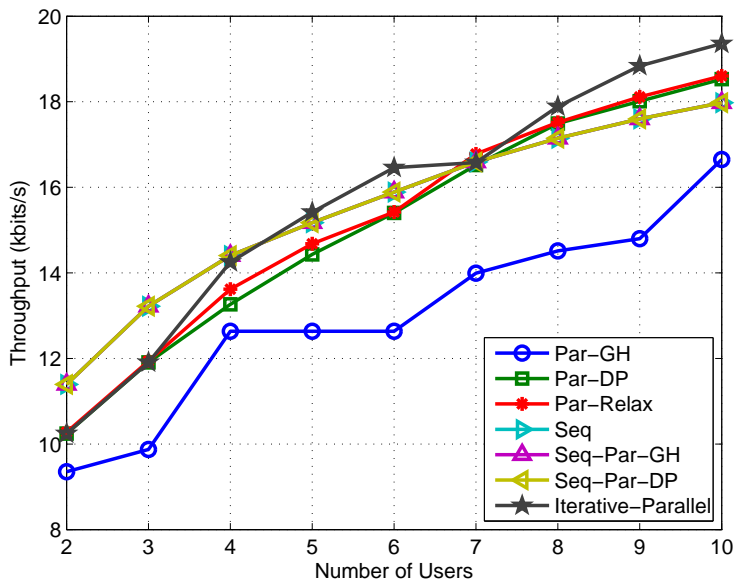}
\caption{Performance comparison for parallel strategy with greedy heuristic (Par-GH), dynamic programming (Par-DP), constraint relaxation (Par-Relax), sequential strategy (Seq), sequential-parallel strategy with greedy heuristic (Seq-Par-GH), sequential-parallel strategy with dynamic programming (Seq-Par-DP), and iterative parallel strategy (Iterative-Parallel) under $OR$ fusion rule versus the number of users. Simulation parameters are $T=5$ ms, $\Gamma=10$ dB, $Q_d=0.9$, $Q_f=0.15$, $M=6$, $\gamma=-5$ dB, $B=0.5f_s=(1,1.5,2,2.5,3,5)$ kHz, and $u=(0.1,0.2,0.3,0.4,0.5,0.3)$ for each channel.}
\label{Fig:S1_05012014_crop}
\end{figure}

In Fig.~\ref{Fig:S1_05012014_crop}, it can be observed that, among the parallel strategies, the constraint relaxation achieves the largest throughput because the solution in equation~(\ref{eq:parallel-objective}) is not necessarily integer hence it provides an upper bound for the parallel strategies. Second, there exists a performance gap between the proposed greedy heuristic strategy and the optimal parallel strategy. This can be explained by the fact that in the heuristic, we only consider the traffic and channel information, but we ignore the impact of the fusion model and thus sensing time. Third, the sequential strategy outperforms the parallel strategies when the number of users is small. This is due to the fact that all channels can be sequentially sensed in a time slot with small sensing time for each channel. In addition, the number of channels is large compared to the number of users so that the sequential strategy would not waste as many spectrum opportunities as the parallel strategies. Finally, as expected, the sequential-parallel strategy performs better than pure sequential and parallel strategies, and the proposed iterative parallel strategy can not outperform the sequential-parallel. The proposed greedy heuristic for the sequential-parallel does not perform well for the selected parameter values. Note that in~Fig.~\ref{Fig:S1_05012014_crop} we consider the~\emph{OR} rule, but for the~\emph{AND} rule we also have similar results.

\subsection{Throughput Comparisons for Parallel and Sequential Strategies}
\begin{figure*}
\centering
\subfigure[Performance comparison versus probability of detection $Q_d$. Simulation parameters are $Q_f=0.15$, $N=3$, $M=4$, $\gamma=-5$ dB, $f_s=2B=5$ kHz, $u=0.3$, $T=5$ ms, and $\Gamma=10$ dB.]{
\includegraphics[width=0.33\columnwidth]{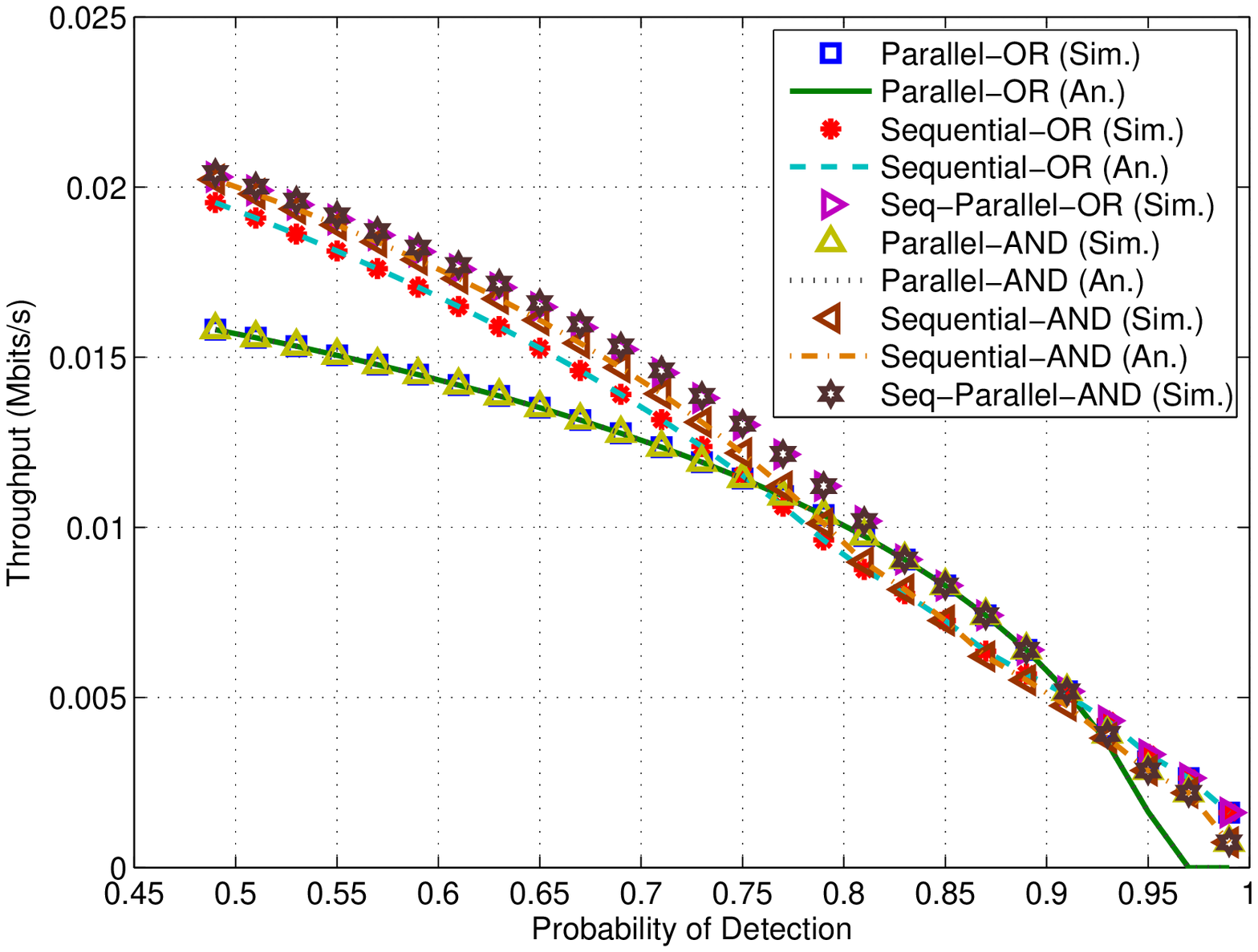}
\label{Fig:S1_Final}
}
\subfigure[Performance comparison versus number of channels. Simulation parameters are $Q_d=0.9$, $Q_f=0.15$, $N=3$, $\gamma=-5$ dB, $f_s=2B=5$ kHz, $u=0.3$, $T=5$ ms, and $\Gamma=10$ dB.]{
\includegraphics[width=0.33\columnwidth]{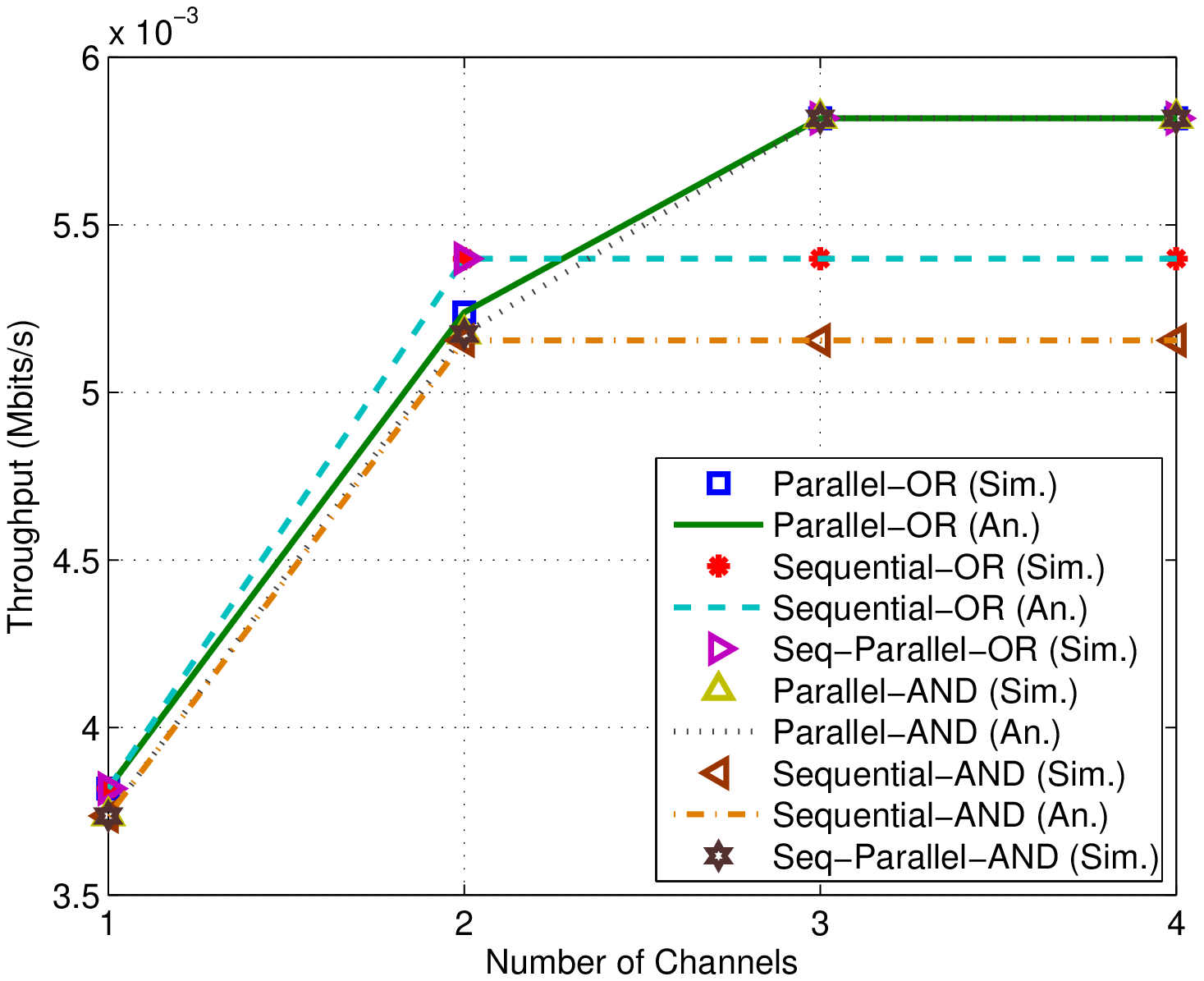}
\label{Fig:S11_Final}
}
\subfigure[Performance comparison versus number of users. Simulation parameters are $Q_d=0.9$, $Q_f=0.15$, $M=4$, $\gamma=-5$ dB, $f_s=2B=5$ kHz, $u=0.3$, $T=5$ ms, and $\Gamma=10$ dB.]{
\includegraphics[width=0.33\columnwidth]{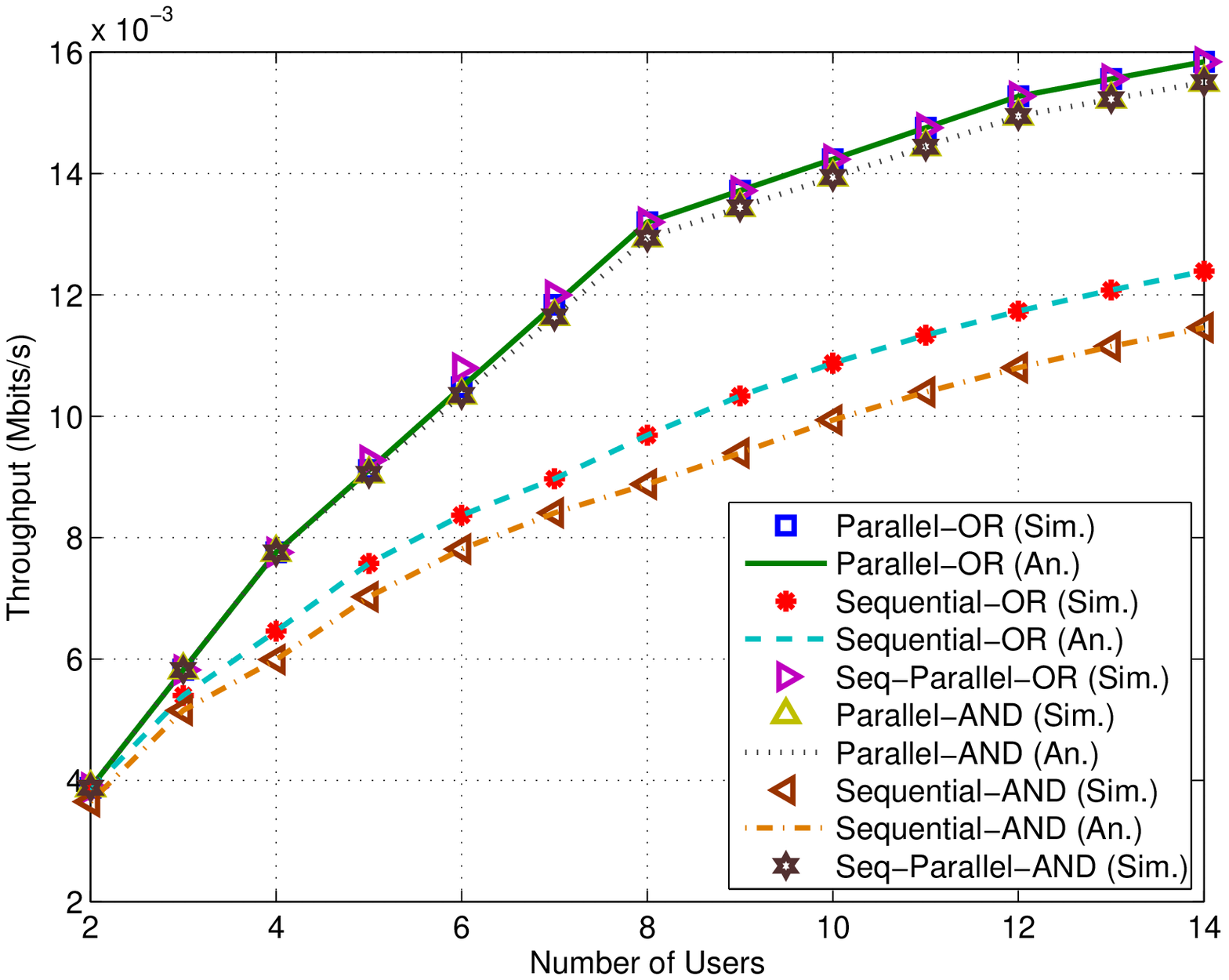}
\label{Fig:S12_Final}
}
\caption{Throughput comparisons for using the~\emph{OR} fusion rule of parallel (Par-OR), sequential (Seq-OR), sequential-parallel (Seq-Par-OR), and the~\emph{AND} fusion rule of parallel (Par-AND), sequential (Seq-AND), sequential-parallel (Seq-Par-AND) strategies with homogeneous channels. Simulation results (Sim.) are shown to be matched with analytical results (An.).}
\label{Fig:throughput_comparisons}
\end{figure*}

To have a better insight on all the strategies, we vary different parameters in Fig.~\ref{Fig:throughput_comparisons} to show their impact on the optimal throughput. We assume that all channels have the same parameters to simplify the results interpretation. 

In Fig.~\ref{Fig:S1_Final}, we first vary various $Q_d$. A higher target $Q_d$ implies a longer sensing time, and we thus observe a performance degradation for all strategies. When the sensing time is short, the sequential strategy guarantees that all channels will be sensed. On the other hand, the parallel strategy senses a maximum of $N$ channels which is less than $M$. Hence it can be seen that the parallel strategy performs better than the sequential when the sensing time is long since the spectrum opportunities in the last channels in the sequential schedule will be very short or null. The sequential-parallel strategy naturally outperforms the sequential and parallel strategies. It can be seen that when the sensing time is very short, the performance of the sequential strategy and the sequential-parallel is the same. This means that the optimal decision by the sequential-parallel strategy also senses all channels by all users. When sensing time is very long, the sequential-parallel results are close to the parallel strategy. 

It can be seen that analytical results provided in Section~\ref{subsec:analytical-discuss} are matched well with the numerical results except for the parallel strategy for very large values of $Q_d$. For the selected simulation parameters, sensing time with a few number of sensors to satisfy a large $Q_d$ is larger than the time slot. We saw that the parallel analytical approach tries to distribute users as much as possible with less cooperation, so the throughput will be zero for large values of $Q_d$. The optimal solution however selects the maximum cooperation on a single channel. 

Comparing the~\emph{AND} and~\emph{OR} rules, the performance is similar for the parallel strategy. Except for very large values of $Q_d$, the parallel strategy, for both~\emph{AND} and~\emph{OR} rules, distributes the users evenly among channels, implying that each user will sense one channel and no cooperation takes place for $N\leq M$. Therefore, the fusion model is irrelevant. For the sequential strategy, we can see that the~\emph{AND} rule outperforms the~\emph{OR} rule when the sensing time is short. It can be explained by the fact that the sensing time is shorter for the~\emph{AND} rule. However for $Q_d=1-Q_f=0.85$, they have the same performance and after that the~\emph{OR} rule performs better since the sensing time becomes shorter for the~\emph{OR} rule. It is worth noting that decreasing $Q_f$ or decreasing the sampling frequency has a similar impact for both rules, as the sensing time increases.

In Fig.~\ref{Fig:S11_Final}, the number of users is fixed to three. When the number of channels $M$ increases, it is expected to have some performance gain. However, the parallel strategy can not sense more than three channels and thus its throughput saturates at three channels. The sequential strategy performance increases with more channels until no more channels can be sensed. Given the sensing time equal to $\tau_N$, maximum $\lfloor\frac{T}{\tau_N}\rfloor$ channels may be sensed in one time slot, which is $\lfloor\frac{5~\text{ms}}{1.8~\text{ms}}\rfloor=2$ channels in this figure. The sequential-parallel strategy may have three user subsets and in each subset, maximum $\lfloor\frac{T}{\tau_1}\rfloor=\lfloor\frac{1~\text{ms}}{3.4~\text{ms}}\rfloor=1$ channels can be sensed. The saturation thus occurs at $M=3$. Discussions related to comparing~\emph{AND} and~\emph{OR} fusion rules are similar to the previous figure. 

When we increase the number of users, it is expected to observe that when $N<M$, the parallel strategy can not sense all channels, so that the sequential strategy outperforms the parallel strategy. When the number of users increases, the parallel strategy is able to sense all channels, so it becomes superior. However in Fig.~\ref{Fig:S12_Final}, the sequential scheme can not sense more than two channels when $N\geq2$, and three channels when $N\geq4$. So its performance improves mostly by the  decrease in the cooperative sensing time. However, the parallel scheme will be able to sense four channels when the number of users increases from $2$ to $4$, and then for larger number of users benefits from cooperative sensing. Its performance is thus always superior to the sequential scheme. 

The other interesting point to observe is that as the number of users increases, the gap between ~\emph{OR} and~\emph{AND} fusion rules increases because cooperative sensing becomes more likely and for the given $Q_d$ and $Q_f$, the \emph{OR} fusion model has shorter cooperative sensing times. 

\subsection{Performance Comparison for Heterogeneous Sensors}
\label{sec:hetsens-numerical}
\begin{figure}%
\centering
\includegraphics[width=0.68\columnwidth]{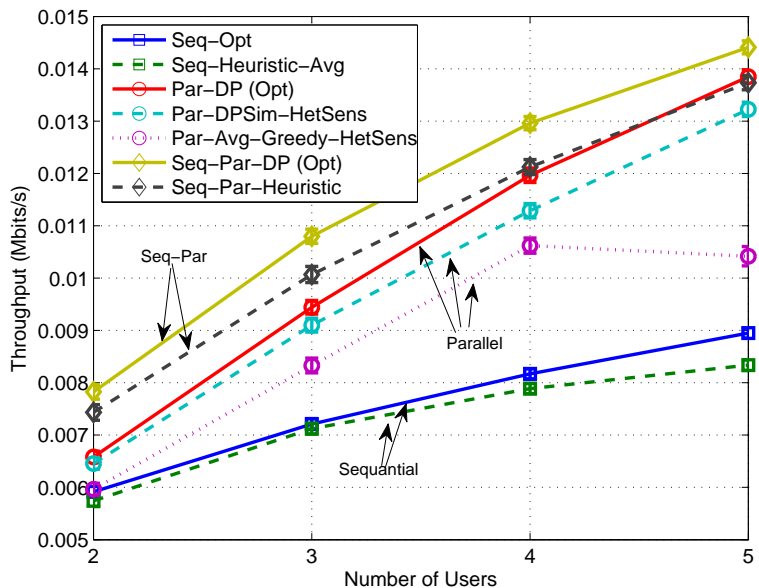}%
\caption{Throughput comparisons for using~\emph{AND} fusion rule for the scenario with heterogeneous sensors. Optimal results and proposed heuristics for Sequential (Seq), Parallel (Par) and Sequential-Parallel (Seq-Par) strategies are compared. The figure is the heterogeneous version of Fig.~\ref{Fig:S12_Final} with the same parameter values except the detection SNR $\gamma$, which is random here. For the noise power, we set its variance $\sigma^2=10^{-8}$ W. The confidence interval for each simulation points is set to be $90\%$.}%
\label{AllschemesMixedMixingN24N5}%
\end{figure}

In this section, we discuss the performance of the proposed solutions for the scenarios with heterogeneous sensors. Due to space limit, we only provide the heterogeneous version of Fig. \ref{Fig:S12_Final} since the results are similar for other scenarios. Further, the results are only for the \emph{AND} fusion rule because with the \emph{OR} rule, mixing any set of SNRs does not worsen the performance and the cooperative sensing time will not be worse than the non-cooperative; therefore, the notion of optimal set of sensors is not applicable. Instead of a fixed detection SNR equal to $-5$ dB, in this section users have an exponentially distributed random SNR with an average of $-5$ dB but limited to the bound $(-15,0)$ dB. Fig.~\ref{AllschemesMixedMixingN24N5} is the heterogeneous version of Fig.~\ref{Fig:S12_Final} comparing sequential, parallel and sequential-parallel strategies together and with the proposed heuristics. We ran the simulation for $500$ runs for all schemes except the sequential-parallel DP for $N=6$, which was repeated 100 times.  However, it can be seen that the confidence intervals are acceptable for $90\%$ confidence. It should be noted that even though the second heuristic proposed for the parallel strategy does not perform well, it is provided to reveal the fact that when two users with different SNRs cooperate, the performance can be worse. For this heuristic, there are four channels, so up to $N=4$, each user has one channel. For $N=5$, two users cooperate, but it may degrade the performance since those cooperating users are not selected optimally in the second heuristic. 
Compared to Fig.~\ref{Fig:S12_Final}, we can also see that throughput is higher here with heterogeneous sensors. This can be explained by Jensen's inequality. Considering throughput as a function of detection SNR, i.e., $R(\gamma)$, we observed that this function is convex in the range of $(-15,0)$ dB (but concave for higher values of SNR). Therefore, it is expected to have $R(\mathbb{E}\{\gamma\}) \leq\mathbb{E}\{R(\gamma)\}$. 
\subsection{Performance Comparison for Primary Traffic and Channel Uncertainty}
\begin{figure}[!t]
\centering
\includegraphics[width=0.68\columnwidth]{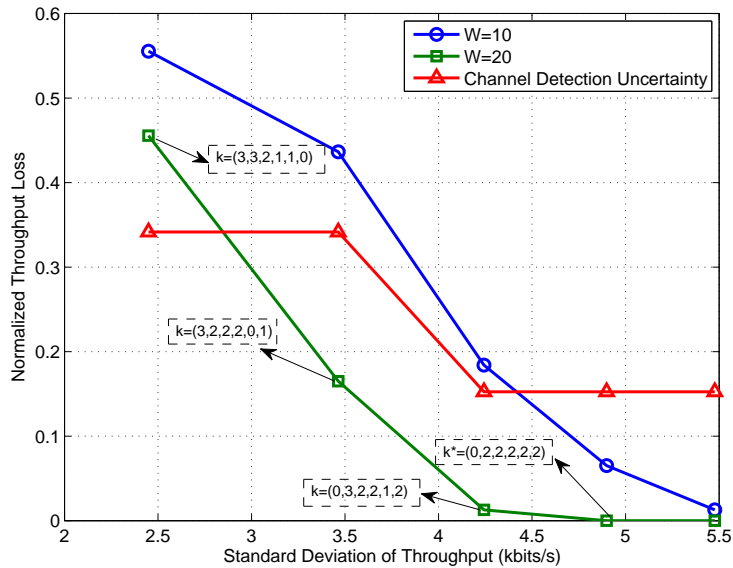}
\caption{Normalized throughput loss comparison for traffic and channel uncertainty with parallel sensing strategy under $OR$ fusion rule versus the standard deviation of throughput. Simulation parameters are $T=5$ ms, $\Gamma=10$ dB, $Q_d=0.9$, $Q_f=0.25$, $N=10$, $M=6$, $\gamma=-5$ dB for the traffic uncertainty, $(\phi_L,\phi_U)=(-15,-1)$ dB, $B=0.5f_s=(1,1.5,2,2.5,3,5)$ kHz, $u=(0.1,0.2,0.3,0.4,0.5,0.3)$ and $P_{00}=0.9$ for each channel.}
\label{Fig:S_R1_30042014_crop}
\end{figure}

In this section, we compare the throughput performance under primary traffic and channel uncertainty by solving~\textit{the Robust Optimization Problem~\ref{optimization:traffic}} and~\textit{Robust Optimization Problem~\ref{optimization:SNR}}, respectively. To demonstrate the results, we choose the proposed parallel sensing strategy as an example with fusion~\textit{OR} rule. For primary traffic uncertainty design in~\textit{Robust Optimization Problem~\ref{optimization:traffic}}, by applying the parallel sensing strategy, the strategy constraint, i.e., $\mathcal{S}\in\mathcal{A}$, would be $\sum\limits_{i=1}^N k_i=N$, and the sensing time $T_I^{(i)}(\mathcal{S})$ is equal to $\tau_{i,k_i}$, as shown in equation~(\ref{eq:parallel-objective}). For channel uncertainty design in~\textit{Robust Optimization Problem~\ref{optimization:SNR}}, by applying the parallel sensing strategy, the strategy constraint is also $\sum\limits_{i=1}^N k_i=N$, while the constant factor is $A^{(i)}(\mathcal{S})=\frac{[Q^{-1}(1-\sqrt[k_i]{1-Q_f})-Q^{-1}(1-\sqrt[k_i]{1-Q_d})]^2}{f_s}$, from equation~(\ref{eq:Coop-sensing_time}). Fig.~\ref{Fig:S_R1_30042014_crop} shows the normalized throughput loss with the standard deviation of the throughput constraint for primary traffic uncertainty case given a constant number of traffic samples $W$ for all channels, and for the channel uncertainty case. Note that they are compared with the same average detection SNR $\gamma$ in $-5$ dB. The normalized throughput loss is defined as the throughput loss from maximal throughput without variance constraint and normalized by it. From the figure, first we observe that, as we increase the standard deviation of the throughput constraint, which means that we have relaxed the robust optimization constraint, we have a lower throughput loss. This is due to the fact that we have more search space for the optimal allocation $k_i$, and hence we can achieve higher throughput. Second, as we increase the number of traffic samples $W$, the throughput loss decreases. This is because as we use more traffic samples to estimate the channel busy rate $u_i$, we have less uncertainty on the estimation. Therefore we can have lower variance for the estimator, which results in lower variance for the throughput. In other words, we can achieve larger throughput with the same variance constraint. Third, we can observe that the decreasing rate for the channel detection uncertainty case is smaller than for the primary traffic uncertainty. This means that if we have channel detection uncertainty, it will be difficult to achieve low throughput loss compared with traffic uncertainty. This is important since in practice, accurately estimating the instantaneous detection SNR can be difficult. Finally, as we increase the constraint, the optimal allocation solution $k_i$ will also change accordingly. Note that as the standard deviation is large enough, the normalized throughput loss will eventually go to zero, i.e., to provide the same optimal allocation as the optimization without constraint, e.g., $\mathbf{k}^*=(0,2,2,2,2,2)$ when $W=20$. In addition, since we have a discrete solution space for $k_i$, if the variance threshold does not increase significantly, the solution may stay the same hence resulting in the ladder type curve as shown for channel detection uncertainty case.
\subsection{Design Examples}
\begin{figure}[!t]
\centering
\includegraphics[width=0.68\columnwidth]{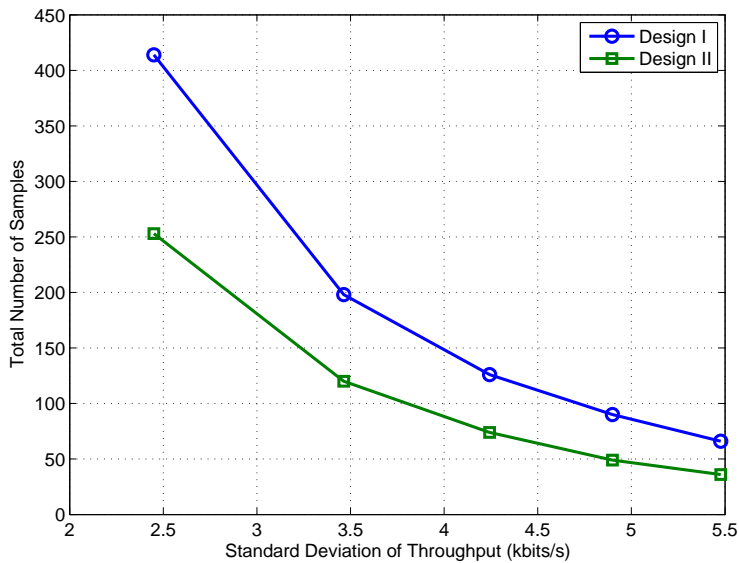}
\caption{Total number of samples comparison for Design I and Design II with parallel sensing strategy under $OR$ fusion rule versus the standard deviation of throughput. Simulation parameters are $T=5$ ms, $\Gamma=10$ dB, $Q_d=0.9$, $Q_f=0.25$, $N=10$, $M=6$, $\gamma=-5$ dB, $B=0.5f_s=(1,1.5,2,2.5,3,5)$ kHz, $u=(0.1,0.2,0.3,0.4,0.5,0.3)$ and $P_{00}=0.9$ for each channel.}
\label{Fig:S_R2_30042014_crop}
\end{figure}

In this section, we elaborate two design examples for primary traffic uncertainty by solving ROP~\ref{optimization:Variance} assuming that we do not have knowledge of primary traffic statistics. To achieve the maximal throughput without variance constraint with the optimal allocation $k^*_i$, we substitute it in ROP~\ref{optimization:Variance} to solve for the minimum number of traffic samples that we need to use to achieve the minimum variance for the throughput. We propose two designs, i.e., homogeneous and heterogeneous estimation, shown as~\textit{Design 1:}~$\min\limits_{W} W,~\mbox{s.t.}~\text{Var}\{\hat R(\mathcal{S})\}\leq\eta$, and~\textit{Design 2:}~$\min\limits_{W_i}\sum\limits_{i=1}^{M}W_i,~\mbox{s.t.}~\text{Var}\{\hat R(\mathcal{S})\}\leq\eta$.
\textit{Design 1} represents the case where we estimate the traffic duty cycle for all channels with the same number of traffic samples, while~\textit{Design 2} is when we estimate the traffic duty cycle on each channel with its own necessary number of traffic samples. Fig.~\ref{Fig:S_R2_30042014_crop} shows the total number of samples needed to achieve a certain standard deviation of throughput. First, as we relax the threshold to have less robustness, we need less number of samples for estimation. Second, since~\textit{Design 2} is a general optimization process considering traffic characteristics for each channel, it can achieve the same standard deviation threshold using less number of samples. Finally, from the system design point of view, for example, if we want to achieve a robust design for the optimal throughput with standard deviation less than $3.55$ kbits/s, we need to use at least $100$ samples for traffic estimation.
\section{Conclusions}
\label{sec:conclusions}
In this paper, we propose and compare several cooperative spectrum sensing strategies for homogeneous and heterogeneous sensors, i.e., sequential, parallel, and sequential-parallel to schedule users to sense multiple channels in order to achieve the optimal throughput. For each strategy, we introduce several solutions including low-complexity heuristic and dynamic programming methods. In addition, we propose a robust scheduling design in terms of both primary traffic and channel detection uncertainty, and a design guideline is also provided for primary traffic estimation given the throughput variation constraint. In terms of throughput performance, we show that with longer sensing time, such as when we have a stringent constraint on probability of detection, smaller number of channels, or larger number of users, the parallel sensing strategy is recommended. Otherwise the sequential sensing strategy should be adopted. A hybrid sequential-parallel sensing strategy has the benefits of both approaches and perform better in almost all scenarios while it suffers from a high complexity limiting its implementation and usage. 

\appendices

\section{Analytical Approximation for the General Parallel Strategy}
\label{apx-relaxation}
\begin{figure}[!t]%
\centering
\includegraphics[width=0.68\columnwidth]{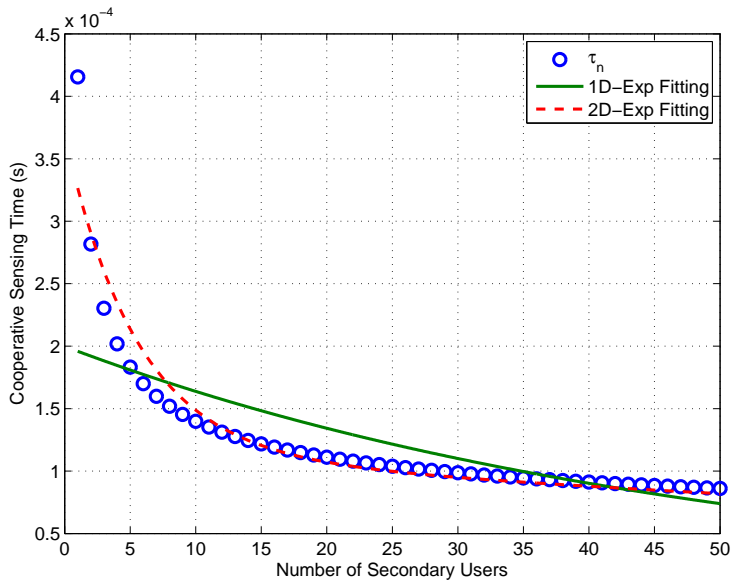}%
\caption{Fitting $\tau_{n}$ with a 1-degree $a e^{-bx}$ and a 2-degree $a e^{-bx}+c e^{-dx}$ exponential functions using~emph{OR} fusion rule with used parameters $Q_d=0.9$, $Q_d=0.1$, $SNR=-5$ dB, and $f_s=5$ kHz. Fitting results are $a \approx 0.0022$ and $b \approx 0.0265$ for 1-D exponential, and $a \approx 0.0030$, $b \approx 0.3088$, $c \approx 0.0015$ and $d \approx 0.0136$ for the 2-D exponential fitting.} %
\label{FittingFigure.pdf}%
\end{figure}
We provided an analytical analysis for the case with homogeneous channels. For a more general case, it can be seen that the convexity of the throughput function of the parallel strategy depends on $\tau_{i,k_i}$ values, so it should be investigated whether the function of sensing time versus the number of collaborators is a convex function or not. For this aim, we define $f(x)=Q^{-1}(\sqrt[k]{1-Q_f})-Q^{-1}(\sqrt[k]{1-Q_d})$ (please see equation~(\ref{eq:Coop-sensing_time}). Examining the Hessian of this function shows that $f(k)$ is not convex unless if there are enough users. The exact threshold for the number of users to have a convex function depends on the other parameters such as $Q_d$ and $Q_f$. 
Therefore, for a large number of users, if we approximate $f(k)$ by an exponential function in the form $ae^{-bk}, a>0, b>0$, the optimal user allocation can be found analytically by convex optimization, which is equal to:
\begin{equation}
k_i^*=\frac{\ln(a b C_i (1-u_i))}{b}-\frac{1}{Mb}{\sum_{j=1}^{M}{a b C_j (1-u_j)}+\frac{bN}{M}},
\label{eq:optimal_allocation}
\end{equation}
where $1\leq i\leq M$. Fig.~\ref{FittingFigure.pdf} confirms that fitting with an exponential function is not accurate when the number of users is low, but is acceptable when there are a large number of users. In practical scenarios, it is expected to have a large number of users, otherwise the problem can easily be solved by a Brute-Force method, so that the exponential approximation and consequently the analytical derivation for optimal assignment are very helpful.
\section{Derivation of the Variance for the Average Estimator}
\label{variance_estimator}
The variance of the proposed $\hat u$ can be written as
\begin{align}
\text{Var}\{\hat u\}
&=\mathbb{E}\{[\hat u-\mathbb{E}\{\hat u\}]^2\}\notag\\
&=\mathbb{E}\left\{\left(\frac{1}{W}\sum\limits_{i=1}^W z_i\right)^2\right\}-u^2\notag\\
&=\frac{1}{W^2}\sum\limits_{i=1}^W\mathbb{E}\left\{z_i^2\right\}+\frac{2}{W^2}\sum\limits_{i=1}^{W-1}\sum\limits_{j=1}^{W-i}\mathbb{E}\left\{z_i z_{i+j}\right\}-u^2.
\label{eq:var1}
\end{align}
The expression $\mathbb{E}\left\{z_i z_{i+j}\right\}$ in equation~(\ref{eq:var1}) is the correlation terms between $z_i$ and $z_{i+j}$, and we denoted as $R_{i,i+j}$. For $R_{i,i+j}$, $\forall j\geq0$, we can solve it by its recursive definition, i.e.,
\begin{align}
R_{i,i+j}
&=\mathbb{E}\{z_i z_{i+j}\}=\Pr\{z_i=1,z_{i+j}=1\}\notag\\
&=\Pr\{z_i=1,z_{i+j-1}=1\}\times P_{11}\notag\\
&+\Pr\{z_i=1,z_{i+j-1}=0\}\times P_{01}\notag\\
&=R_{i,i+j-1}\times P_{11}+(u-R_{i,i+j-1})\times P_{01}\notag\\
&=R_{i,i+j-1}\times r+u\times P_{01},
\label{eq:correlation}
\end{align}
where $r=P_{11}-P_{01}$. The initial condition for the recursive equation~(\ref{eq:correlation}) is $R_{i,i}=\mathbb{E}\{z_i^2\}=u$. Hence, solving the recursive equation~(\ref{eq:correlation}) gives the result
\begin{align}
R_{i,i+j}=\frac{uP_{01}(1-r^j)}{1-r}+u r^j.
\label{eq:correlation_final} 
\end{align}
By plugging equation~(\ref{eq:correlation_final}) into equation~(\ref{eq:var1}), we can simplify it as the result shown in equation~(\ref{Var_u}).
\section{Proof of Lemma~\ref{lemma:equivalence}}
\label{proof_of_equivalence}
First consider ROP~$2$. Define the objective function as $f(\mathbf{W})=\sum\limits_{i=1}^{M}W_i$, and the constraint function as $g(\mathbf{W})=\text{Var}\{\hat R(\mathcal{S})\}$. Assume we obtain the optimal solution $\mathbf{W}^*=(W_1^{*},W_2^{*},\cdots,W_M^{*})$ to minimize the function $f(\mathbf{W})$. Define $\epsilon=f(\mathbf{W}^{*})$. This means for any other feasible solutions $\mathbf{W}\neq\mathbf{W}^*$ satisfying $g(\mathbf{W})\leq\eta$, we always have $f(\mathbf{W})\geq\epsilon$. The above statement is equivalent to if we find a solution $\mathbf{W}$ such that $f(\mathbf{W})\leq\epsilon$, we should always have the constraint being violated, i.e., $g(\mathbf{W})\geq\eta$. Among all of these possible $\mathbf{W's}$, we want to find the one with minimum $g(\mathbf{W})$, which forms the proposed ROP~3. We can see that since all the possible $\mathbf{W}\neq\mathbf{W}^*$ gives $g(\mathbf{W})\geq\eta$, while we know $g(\mathbf{W}^*)\leq\eta$ by the original constraint, we can conclude that $\mathbf{W}^*$ is still the optimal solution for ROP~$3$ to minimize $g(\mathbf{W})$. Note that we can also proof the other equivalence by the same technique, i.e., given the optimal solution for ROP~$3$, we can show that it is also the optimal solution for ROP~$2$.
\bibliographystyle{IEEEtran}
\bibliography{IEEEabrv,bibglobe}

\begin{thebibliography}{10}
\providecommand{\url}[1]{#1}
\csname url@samestyle\endcsname
\providecommand{\newblock}{\relax}
\providecommand{\bibinfo}[2]{#2}
\providecommand{\BIBentrySTDinterwordspacing}{\spaceskip=0pt\relax}
\providecommand{\BIBentryALTinterwordstretchfactor}{4}
\providecommand{\BIBentryALTinterwordspacing}{\spaceskip=\fontdimen2\font plus
\BIBentryALTinterwordstretchfactor\fontdimen3\font minus
  \fontdimen4\font\relax}
\providecommand{\BIBforeignlanguage}[2]{{%
\expandafter\ifx\csname l@#1\endcsname\relax
\typeout{** WARNING: IEEEtran.bst: No hyphenation pattern has been}%
\typeout{** loaded for the language `#1'. Using the pattern for}%
\typeout{** the default language instead.}%
\else
\language=\csname l@#1\endcsname
\fi
#2}}
\providecommand{\BIBdecl}{\relax}
\BIBdecl

\bibitem{Arash14}
A.~{Azarfar}, C.-H. {Liu}, J.-F. {Frigon}, B.~{Sanso}, and D.~{Cabric},
  ``Cooperative spectrum sensing scheduling optimization in multi-channel
  dynamic spectrum access networks,'' Dec. 2014, accepted in IEEE GLOBECOM.

\bibitem{zhao07}
Q.~Zhao and B.~Sadler, ``A survey of dynamic spectrum access,'' \emph{{IEEE}
  Signal Processing Magazine}, vol.~24, no.~3, pp. 79--89, May 2007.

\bibitem{Hoang08}
Y.-C. {Liang}, Y.~{Zeng}, E.~C.~Y. {Peh}, and A.~T. {Hoang},
  ``Sensing-throughput tradeoff for cognitive radio networks,'' \emph{{IEEE}
  Trans. Wireless Commun.}, vol.~7, no.~4, pp. 1326--1337, Apr. 2008.

\bibitem{Brodersen06}
D.~{Cabric}, A.~{Tkachenko}, and R.~W. {Brodersen}, ``Spectrum sensing
  measurements of pilot, energy, and collaborative detection,'' in \emph{IEEE
  Military Commun. Conf. (MILCOM)}, Oct. 2006.

\bibitem{Shin08}
H.~{Kim} and K.~G. {Shin}, ``Efficient discovery of spectrum opportunities with
  mac-layer sensing in cognitive radio networks,'' \emph{{IEEE} Trans. Mobile
  Comput.}, vol.~7, no.~5, pp. 553--545, May 2008.

\bibitem{liu13}
C.-H. Liu, J.~Tran, P.~Pawelczak, and D.~Cabric, ``Traffic-aware channel
  sensing order in dynamic spectrum access networks,'' \emph{IEEE JSAC},
  vol.~31, no.~11, pp. 2312--2323, Nov. 2013.

\bibitem{azarfar13}
A.~Azarfar, J.~Frigon, and B.~Sanso, ``User-differentiated channel recovery in
  multi-channel cognitive radio networks,'' in \emph{International Conference
  on DRCN}, March 2013, pp. 242--249.

\bibitem{Datla09}
D.~Datla, R.~Rajbanshi, A.~M. Wyglinski, and G.~Minden, ``An adaptive spectrum
  sensing architecture for dynamic spectrum access networks,'' \emph{{IEEE}
  Transactions on Wireless Communications}, vol.~8, no.~8, pp. 4211--4219,
  August 2009.

\bibitem{Giannakis08}
A.~{Ribeiro} and G.~B. {Giannakis}, ``Robust stochastic routing and scheduling
  for wireless ad-hoc networks,'' in \emph{IEEE Conference on Wireless
  Communications and Mobile Computing}, Aug. 2008.

\bibitem{Cui14}
Q.~{Liu}, X.~{Wang}, and Y.~{Cui}, ``Robust and adpative scheduling of
  sequential periodic sensing for cognitive radios,'' \emph{{IEEE} J. Select.
  Areas Commun.}, vol.~32, no.~3, pp. 503--515, Mar. 2014.

\bibitem{Digham12}
L.~{Hesham}, A.~{Sultan}, M.~{Nafie}, and F.~{Digham}, ``Distributed spectrum
  sensing with sequential ordered transmissions to a cognitive fusion center,''
  \emph{{IEEE} Trans. Signal Processing}, vol.~60, no.~5, pp. 2524--2538, May
  2012.

\bibitem{Casadevall13}
M.~{Lopez-Benitez} and F.~{Casadevall}, ``Time-dimension models of spectrum
  usage for the analysis, design and simulation of cognitive radio networks,''
  \emph{IEEE TVT}, vol.~62, no.~5, pp. 2091--2104, Jun. 2013.

\bibitem{ghasemi05}
A.~Ghasemi and E.~Sousa, ``Collaborative spectrum sensing for opportunistic
  access in fading environments,'' in \emph{IEEE DySPAN}, Nov 2005, pp.
  131--136.

\bibitem{bertsekas05}
D.~P. Bertsekas, \emph{Dynamic Programming and Optimal Control}.\hskip 1em plus
  0.5em minus 0.4em\relax Athena Scientific, 2005, vol. I, 3rd Edition.

\bibitem{martello90}
S.~Martello and P.~Toth, \emph{Knapsack problems: Algorithms and computer
  interpretations}.\hskip 1em plus 0.5em minus 0.4em\relax Wiley-Interscience,
  1990.

\bibitem{Cabric13}
W.~{Gabran}, C.-H. {Liu}, P.~{Pawe{\l}czak}, and D.~{Cabric}, ``Primary user
  traffic estimation for dynamic spectrum access,'' \emph{{IEEE} J. Select.
  Areas Commun.}, vol.~31, no.~3, pp. 544--558, Mar. 2013.

\end{thebibliography}

\vspace{-1cm}
%
\begin{IEEEbiographynophoto}{Chun-Hao~Liu} received the BS and MS degrees in electronics engineering from National Chiao Tung University and National Taiwan University, in 2007 and 2009, respectively. He is currently working toward the PhD degree in the Department of Electrical Engineering, University of California, Los Angeles.
His research interests are signal processing and algorithms design for wireless communications. He is working on development of algorithms and analysis for cognitive radio networks.
\end{IEEEbiographynophoto}
\vspace{-0.9cm}
\begin{IEEEbiographynophoto}{Arash Azarfar}
received the B.Sc. degree in 2003 and the M.Sc. degree in 2005
both in computer engineering from Sharif University
of Technology, Tehran, Iran, and the Ph.D.
degree in Electrical Engineering from {\'E}cole Polytechnique de Montr{\'e}al in 2014.
His research interests include quality of service, reliability and performance evaluation in cognitive radio and multimedia communication networks.
\end{IEEEbiographynophoto}
\vspace{-0.9cm}
\begin{IEEEbiographynophoto}{Jean-Fran\c{c}ois Frigon}
received the
B.Eng. degree from {\'E}cole Polytechnique de Montr{\'e}al,
Montr{\'e}al , QC, Canada, in 1996, the M.A.Sc.
degree from the University of British Columbia,
Vancouver, BC, Canada, in 1998, and the Ph.D.
degree from the University of California at Los
Angeles (UCLA), in 2004.
He joined the Electrical
Engineering department at {\'E}cole Polytechnique de Montr{\'e}al in 2004 where he
is currently a Full Professor. His research interests include wireless networks,
MAC protocols, cognitive network  and multiple antennas systems.
\end{IEEEbiographynophoto}
\vspace{-0.9cm}
\begin{IEEEbiographynophoto}{Brunilde Sans\`o}
is a full professor of electrical engineering
at {\'E}cole Polytechnique de Montr{\'e}al and director of the LORLAB. Her
interests are in performance, reliability, design, and optimization of wireless and
wireline networks. She is a recipient of several awards, Associate Editor of
Telecommunication Systems, and editor of two books on planning and performance.
\end{IEEEbiographynophoto}
\vspace{-0.9cm}
\begin{IEEEbiographynophoto}{Danijela~Cabric} (S'96--M'07) received the Dipl. Ing. degree from the University of Belgrade, Serbia, in 1998, and the M.Sc. degree in electrical engineering from the University of California, Los Angeles, in 2001. She received her Ph.D. degree in electrical engineering from the University of California, Berkeley, in 2007, where she was a member of the Berkeley Wireless Research Center. In 2008, she joined the faculty of the Electrical Engineering Department at the University of California, Los Angeles as an Assistant Professor. Dr. Cabric received the Samueli Fellowship in 2008, the Okawa Foundation Research Grant in 2009, Hellman Fellowship in 2012 and the National Science Foundation Faculty Early Career Development (CAREER) Award in 2012. She serves as an Associate Editor in IEEE Journal on Selected Areas in Communications (Cognitive Radio series) and IEEE Communications Letters, and TPC Co-Chair of 8th International Conference on Cognitive Radio Oriented Wireless Networks (CROWNCOM) 2013. Her research interests include cognitive radio systems and spectrum sensing, VLSI architectures of signal processing and digital communication algorithms, and their performance analysis and experiments on embedded system platforms.
\end{IEEEbiographynophoto}
\vfill

\end{document}